%% file: neurips_2026.tex
\documentclass{article}

 \usepackage[eandd, final]{neurips_2026}

\usepackage[utf8]{inputenc} 
\usepackage[T1]{fontenc}    
\usepackage{hyperref}       
\usepackage{url}            
\usepackage{booktabs}       
\usepackage{amsfonts}       
\usepackage{nicefrac}       
\usepackage{microtype}      
\usepackage[table]{xcolor}  
\usepackage{amsmath}
\usepackage{graphicx}
\usepackage{placeins}
\usepackage{listings}
\usepackage{tikz}
\usetikzlibrary{positioning}
\usepackage{fontawesome5}

\title{SCDBench: A Benchmark for LLM-Based Smart Contract Decompilers}

\author{%
  Kaihua Qin \\
  University of Warwick
  \And
  Dawn Song \\
  UC Berkeley
  \And
  Arthur Gervais\\
  University College London
}

\begin{document}

\maketitle

\begin{abstract}
Smart contract decompilation aims to recover high-level source code from bytecode, but evaluating decompilers remains difficult because existing studies use narrow datasets, inconsistent metrics, and limited semantic consistency checks. This gap is increasingly important as large language models (LLMs) begin to generate source-like Solidity that may compile and appear plausible, even when its semantics diverge from the original contract. We introduce SCDBench, a dataset and benchmark methodology for LLM-based smart contract decompilation. The dataset contains~$600$ real-world Solidity contracts with paired bytecode inputs, ground-truth source code, and replayable semantic checkpoints. SCDBench evaluates decompiler outputs through four cumulative stages: format completeness, compilability, Application Binary Interface (ABI) recovery, and semantic consistency via differential replay. We evaluate Claude Opus 4.7, GPT-5.3-Codex, and GLM-5 in a zero-shot decompilation setting, including GLM-5 variants with and without extended reasoning and a zero-shot compilation-repair setting. The results show that frontier LLMs can often produce structured and compilable Solidity, but achieving semantic consistency remains far from solved: the best-performing frontier model perfectly decompiles only~$42/600$ contracts. We further show that introducing same-model compilation repair substantially improves performance at modest additional cost. SCDBench establishes a common ground for rigorous, reproducible evaluation and aims to accelerate the development of reliable smart contract decompilers for blockchain security and transparency.
\end{abstract}

\section{Introduction}
Smart contracts~\cite{wood2014ethereum} are programs deployed on blockchains that manage digital assets, enforce agreements, and underpin decentralized applications. They have transformed the blockchain ecosystem from simple value transfer to a programmable financial and computational infrastructure. Today, smart contracts secure over~$160$B USD across decentralized finance, non-fungible tokens, and governance systems.\footnote{\url{https://defillama.com/}}

While the bytecode of smart contracts is publicly accessible on-chain, their source code is not always made available. Only a small fraction of contracts are voluntarily verified by developers and published in high-level languages such as Solidity. At the time of writing, more than~$99\%$ of the smart contracts on Ethereum, the largest smart contract-enabled blockchain, are unverified, leaving only low-level bytecode accessible.\footnote{\url{https://etherscan.io/contractsverified}} This opacity hinders transparency, auditability, and accountability, as developers, users, and security researchers cannot easily understand or validate the semantics of deployed contracts. Recovering high-level representations from bytecode, i.e., decompilation, is therefore essential for security analysis and systematic understanding of the blockchain ecosystem.

Traditional smart contract decompilers~\cite{zhou2018erays,grech2019gigahorse,grech2022elipmoc,lagouvardos2025incredible} have relied on program analysis techniques, such as control-flow reconstruction and type inference. These tools often stop at producing structured intermediate representations (e.g., annotated pseudo-code) that are easier to follow than raw bytecode, but still challenging for developers to interpret. More recently, with the rise of large language models (LLMs), LLM-based decompilers have emerged. LLMs can greatly enhance readability and even generate source-like output that compiles back to bytecode, enabling correctness to be validated automatically. However, the field lacks a standardized benchmark. Existing tools are typically evaluated on proprietary or narrowly scoped datasets, with inconsistent metrics, making it difficult to compare methods, reproduce results, or assess real-world effectiveness. As a result, there is no clear understanding of the relative strengths and limitations of different approaches.

In this paper, we introduce SCDBench, a dataset and benchmark methodology for smart contract decompilation. Our contributions are threefold:

\textbf{Dataset.} We construct a compact but diverse benchmark dataset with~$600$ real-world Solidity contracts sampled from verified Ethereum deployments. The dataset is built from~$772{,}736$ exact-bytecode-unique contracts, filters near-duplicate template variants, and stratifies examples into easy, medium, and hard complexity strata. Each benchmark instance contains the bytecode input and the corresponding Solidity ground truth. To make semantic consistency evaluation reproducible, we also generate~$227{,}383$ concrete semantic test cases covering~$14{,}553$ public functions.

\textbf{Benchmarking methodology.} We define a staged evaluation protocol that measures decompilation quality beyond readability. The pipeline checks format completeness, Solidity compilability, ABI recovery, and semantic consistency under replayed test cases. This methodology provides a reusable evaluation framework for comparing future smart contract decompilers under the same contracts, prompts, and semantic checkpoints.

\textbf{Evaluations and insights.} We benchmark frontier closed-model baselines, represented by Claude Opus 4.7 and GPT-5.3-Codex, and an open-weight baseline, represented by GLM-5. Our evaluations provide three main insights for future decompiler design: frontier and reasoning-capable models are substantially stronger, compilation repair is a simple but effective step toward agentic decompilation, and achieving semantic consistency remains the central open challenge.

By establishing a common ground for evaluation, our benchmark aims to accelerate progress in smart contract decompilation. We believe it will foster reproducibility, enable fair comparison, and ultimately drive the development of more reliable tools for blockchain security and transparency.
The benchmark \href{https://doi.org/10.7910/DVN/IPPTIX}{dataset} and \href{https://github.com/KaihuaQin/SCDBench}{evaluation artifacts}, including prompts, scripts, and model outputs, are publicly available to support reproduction and future comparison.

\section{Related Work}
Smart contract decompilation lifts EVM bytecode into more comprehensible high-level representations. Erays~\cite{zhou2018erays} reconstructs control flow and lifts stack operations into a register-based form to produce pseudocode. Gigahorse~\cite{grech2019gigahorse} uses Datalog-based analysis to recover a three-address intermediate representation, while Elipmoc~\cite{grech2022elipmoc} adds transactional context sensitivity and path-sensitive function reconstruction. Shrnkr~\cite{lagouvardos2025incredible} improves scalability and precision through shrinking context sensitivity and control-flow normalization. Other tools include the pattern-matching-based Panoramix\footnote{\url{https://github.com/eveem-org/panoramix}} and the symbolic-execution-based Heimdall-rs.\footnote{\url{https://github.com/Jon-Becker/heimdall-rs}} \citet{liu2023empirical} compare five decompilers in terms of success rates, performance, ABI recovery, and resilience to compiler optimizations. LLM-based approaches include fine-tuning on contracts lifted into structured three-address code~\cite{david2025decompiling} and DiSCo's frozen-LLM pipeline, which combines semantic-unit representations, graph-based variable naming, and prompt synthesis~\cite{su2025disco}. Despite these advances, decompiler evaluations have largely emphasized syntactic and structural recovery using heterogeneous datasets and protocols. DiSCo is close in spirit to SCDBench through its explicit evaluation metrics, but limited transparency in dataset construction and evaluation protocols hinders reproducibility and systematic comparison. SolEval~\cite{peng2025soleval} benchmarks the related task of repository-level Solidity generation rather than decompilation from deployed bytecode. SCDBench addresses this gap with a shared dataset and staged evaluation of compilability, interface recovery, and semantic consistency.

SCDBench's semantic test generation draws on techniques developed for smart-contract fuzzers~\cite{jiang2018contractfuzzer,wustholz2020harvey,choi2021smartian,shou2023ityfuzz}, including ABI-guided input generation and stateful call sequences. Unlike vulnerability-oriented fuzzing and symbolic execution~\cite{so2021smartest}, SCDBench checks behavioral consistency between original and decompiled contracts by comparing returns, reverts, logs, and storage changes under matched execution conditions. We compare the code coverage of SCDBench and ItyFuzz in Appendix~\ref{app:cr-fuzzing}.

\section{Dataset}
Although fewer than~$1\%$ of Ethereum contracts are source-verified, this subset still contains hundreds of thousands of public Solidity programs. These contracts provide a useful basis for benchmarking: they cover diverse application domains, coding styles, compiler versions, optimization settings, and complexity levels, while reflecting code actually deployed on chain.

In this work, we construct a benchmark dataset with~$600$ real-world Solidity contracts sampled from these verified Ethereum deployments. Each contract defines a bytecode-to-source decompilation task: the input is EVM bytecode, and the ground truth is the corresponding Solidity source code. The benchmark is split into three equal complexity strata: easy, medium, and hard, with~$200$ contracts in each stratum. In addition to bytecode and ground-truth source code, we also generate test cases for every public function to support semantic consistency checking, a critical step in our benchmark pipeline (Section~\ref{sec:metrics}). Across the~$600$ contracts, which expose~$14{,}553$ public functions in total, we generate~$227{,}383$ concrete test cases ($15.6$ test cases per public function on average). In the following, we detail our methodology for creating this benchmark dataset.

\subsection{Design Principles}
We design the dataset to be compact enough for expensive model-based evaluation while still covering diverse real-world contracts. This prioritizes the following three principles.

\textbf{Real-world grounding.}
The dataset should consist of contracts that were actually deployed on Ethereum,\footnote{We focus on Ethereum because it is the largest smart contract blockchain by application ecosystem, and its verified contracts provide the most diverse public source corpus for this benchmark.} not synthetic programs or toy examples. This keeps the decompilation tasks close to the code encountered by users, auditors, and security researchers.

\textbf{Controlled diversity.}
The dataset should measure decompilation ability across compiler generations, contract families, bytecode sizes, interface sizes, and state-interaction patterns. At the same time, common templates and mass-generated variants should not dominate the evaluation.

\textbf{Reproducible semantic consistency evaluation.}
To support our benchmark methodology, the dataset should enable semantic consistency evaluation. Public-function-level inputs and semantic checkpoints should be reusable for any decompiler output, and the generated tests should achieve high code coverage so that semantic consistency results are meaningful.

\subsection{Dataset Construction}
\textbf{Corpus.}
We crawled all verified Solidity smart contracts on Ethereum deployed before January~1, 2026. After removing exact bytecode duplicates, the source corpus contains~$772{,}736$ unique bytecode and source pairs. This gives a large pool of realistic decompilation tasks while avoiding literal duplicate deployments. However, the corpus is still highly skewed: Solidity~v0.8 accounts for~$77.45\%$ of rows, and ERC20-like contracts account for~$63.46\%$. This skew reflects real deployment practice, but direct sampling would allow a few common templates to dominate a compact benchmark.

\textbf{Near-duplicate filtering.}
We intend to construct the benchmark dataset in a diverse manner, covering different compiler settings, bytecode lengths, interface sizes, control-flow complexity, and state-interaction patterns. To do so, we avoid minor template variants that differ only in constants, addresses, names, or small wrapper edits. We therefore apply near-duplicate filtering before sampling, retaining at most one representative from each near-duplicate group. The goal is not to infer semantic classes, but to avoid spending benchmark budget on near-identical decompilation tasks. We present the details of near-duplicate filtering in Appendix~\ref{app:near-duplicate-filtering}.

\textbf{Sampling.}
To understand model performance across contracts of varying complexity, we define three complexity strata based on bytecode size, interface size, control flow, storage operations, external calls, and low-level Solidity features. These proxies are motivated by traditional program-analysis and decompiler evaluation, and it is not obvious a priori that they also predict LLM decompilation difficulty. We therefore use them as interpretable stratification axes rather than as a ground-truth hardness measure. We present the detailed scoring rule in Appendix~\ref{app:difficulty-scoring}. From the filtered corpus, we sample~$200$ contracts from each complexity stratum, for~$600$ contracts in total. Although~$600$ contracts is not large compared with the source corpus, it is a deliberate balance between diversity and evaluation cost: frontier-model decompilation, repair, compilation, ABI extraction, and semantic replay must all be run on the same contracts for paired comparison. We present the detailed sampling rules in Appendix~\ref{app:sampling}. The resulting sample covers all Solidity compiler minor versions from~v0.4 to~v0.8, and the median bytecode size increases from~$2{,}924$ bytes in the easy split to~$15{,}342$ bytes in the hard split (details are presented in Appendix~\ref{app:sample-characteristics}). We examine the relationship between contract complexity and observed model performance in Section~\ref{sec:cr-validation}.

\textbf{External validity.}
SCDBench uses source-verified contracts, which may differ from the broader population of deployed Ethereum contracts. To assess this difference, we randomly sampled~$1{,}000{,}000$ deployed contracts and compared their bytecode with the verified corpus after removing duplicates. The deduplicated sample of deployed contracts has a smaller median bytecode size than the source-verified corpus ($2{,}100$ versus~$6{,}831$ bytes). We find that standard minimal proxies, tiny contracts with minimal logic beyond delegating execution to another contract, account for~$16.94\%$ of the deduplicated sample but are absent from our source-verified corpus. We therefore emphasize that SCDBench targets diversity within source-verified Solidity rather than representing all deployed contracts. The detailed comparisons are presented in Appendix~\ref{app:cr-external-validity}.

\subsection{Semantic Test Cases}
\label{sec:test-cases}
The benchmark dataset includes executable semantic checkpoints for evaluating whether decompiled source code remains semantically consistent with the original contract. We package test cases directly with the dataset, so decompilers can be directly evaluated against the same semantic evidence. Our approach is similar in spirit to differential fuzzing~\cite{yang2021finding}: we execute the ground-truth contract on concrete inputs, record the observed execution outcomes, and later replay the same inputs on the recompiled decompiler output. A replay succeeds only when the decompiler output matches the recorded return data or revert status, emitted logs, and target-contract storage changes for state-changing functions. Test cases are organized by function, which allows semantic comparison at the function level rather than requiring the entire contract interface to be recovered first.

At a high level, the generator combines deterministic random inputs, literals and constants extracted from the contract, boundary values for common guards, and short stateful setup sequences for functions that require prior state. Each case starts from the same freshly deployed target state on a fixed mainnet fork, with state reset between cases. Stateful cases add one setup call before the measured call. Appendix~\ref{app:test-cases} gives the detailed construction and replay format.

For the~$600$ smart contracts, we generate~$227{,}383$ concrete cases covering~$14{,}553$ public functions. To assess how thoroughly contract code is exercised during testing, we measure code coverage and obtain~$77.8\%$ for the benchmark. Note that while higher-coverage tests may reveal additional semantic mismatches, passing these checkpoints does not prove equivalence over all executions. Smart contract fuzzers~\cite{jiang2018contractfuzzer,wustholz2020harvey,choi2021smartian,shou2023ityfuzz,wu2024we} and symbolic execution~\cite{so2021smartest,xia2026symgpt}, with appropriate modifications, may further increase coverage, which we leave to future work.


\section{Benchmark and Evaluation}

SCDBench evaluates source-level decompilation from EVM bytecode to Solidity. For each contract, a decompiler receives the bytecode representation and may use non-unique public signature hints when they are available. The expected output is a self-contained Solidity implementation that can be compiled, exposes the correct external interface, and preserves the semantics of recovered functions.

\subsection{Metrics}\label{sec:metrics}
Evaluating smart contract decompilers requires both syntactic and semantic consistency checks. We define four stages, ordered from basic output validity to semantic consistency.

\textbf{Format completeness.}
This metric checks whether the decompiler output contains all fields needed for evaluation: compiler settings, self-contained Solidity code, and an unambiguous target contract name. It measures whether the output is structurally complete, not whether the code is correct.

\textbf{Compilability.}
This metric checks whether the recovered Solidity is executable as a real program. A contract passes only if the generated code compiles successfully and produces a valid target artifact. It separates compilable Solidity from source-like pseudocode that may look plausible but cannot support downstream checks.

\textbf{ABI recovery.}
This metric evaluates whether the decompiler reconstructs the public interface. We compare recovered public function signatures against the ground-truth ABI and report precision, recall, and F1. Extra generated functions count as false positives, while missing ground-truth functions count as false negatives.

\textbf{Semantic consistency.}
This metric checks whether recovered functions remain semantically consistent with the original bytecode. For each recovered public function, we replay the fixed test cases described in Section~\ref{sec:test-cases} and compare them against ground-truth checkpoints. A function is considered semantically consistent only if all replayed cases match, including return data, revert status, emitted logs, and touched storage-slot changes for state-changing functions. This stage is intentionally strict: passing compilation and ABI recovery does not imply semantic consistency.

Together, these stages form a progressive evaluation pipeline. Format completeness checks whether the output is usable, compilability checks whether it is executable, ABI recovery checks whether the interface is correct, and semantic consistency checks whether recovered execution outcomes match the original contract.

We use cumulative scoring for the staged metrics: results are reported on the original benchmark denominator, not only on contracts or functions that survive previous stages. If a contract or function fails an earlier stage, it is counted as failed in every later applicable stage. Thus, a non-compiling output contributes no recovered ABI entries, and a public function missing from ABI recovery is counted as not semantically consistent. This convention prevents later metrics from being inflated by excluding outputs that fail earlier stages. We also report stage-level metrics restricted to outputs that reach each stage (Appendix~\ref{app:cr-diagnostics}).

\subsection{Evaluation Setup}\label{sec:evaluation-setup}
\textbf{Zero-shot decompilation.}
We evaluate models in a zero-shot decompilation setting. Each model receives the full contract bytecode rendered as EVM assembly and is asked to produce self-contained Solidity code. The assembly is an exact textual rendering of the bytecode: it preserves the opcode sequence while presenting operations more clearly than a raw hexadecimal string. When available, the input also includes public function-selector and event-signature candidate mappings. In the EVM ABI, each public function is called through a four-byte selector, computed as the first four bytes of the hash of its function signature, e.g., \texttt{transfer(address,uint256)} maps to \texttt{0xa9059cbb}. Mappings from common selectors to candidate signatures are publicly available,\footnote{\url{https://4byte.sourcify.dev/}} but they are non-unique and uncertain: one selector may correspond to multiple possible signatures, and the model must decide which hints to use. These mappings supplement the selectors and event topics already present in the assembly and do not provide ground-truth source. We discuss the impact of these mappings on decompilation performance in Section~\ref{sec:cr-validation}.

\begin{figure}[t]
\centering
\begin{tikzpicture}[
  node distance=0.3cm,
  every node/.style={
    rounded corners=2pt,
    font=\footnotesize\ttfamily,
    draw,
    align=left,
    text width=\dimexpr\linewidth-12.4pt\relax,
    minimum width=0pt,
    inner sep=6pt
  }
]

\node[fill=gray!10] (system) {{\faServer}\ System:\\
You are a specialized engine for decompiling EVM assembly into Solidity. Respond with exactly three XML-style blocks in this order: <compiler>, <code>, and <contract\_name>. The <compiler> block is one JSON object, e.g., \{"version":"v0.8.20","settings":\{...\}\}. The <code> block must be complete, self-contained Solidity with no imports.};

\node[fill=blue!5, below=of system] (human) {{\faUser}\ Human:\\
\texttt{Decompile the following EVM assembly into Solidity.\textbackslash n\textbackslash n Assembly:\textbackslash n <assembly>\textbackslash n PUSH1 0x80 PUSH1 0x40 MSTORE ...\textbackslash n </assembly>}};

\node[fill=green!5, below=of human] (ai) {{\faRobot}\ AI:\\
\texttt{<compiler>\{\textbackslash"version\textbackslash": \textbackslash"v0.8.20\textbackslash", \textbackslash"settings\textbackslash":  ...\}</compiler>} \\
\texttt{<code>pragma solidity \^{}0.8.20;\textbackslash n contract Decompiled \{\textbackslash n\qquad function withdraw() external \{...\}\textbackslash n\}</code>} \\
\texttt{<contract\_name>Decompiled</contract\_name>}};
\end{tikzpicture}
\caption{Zero-shot decompilation prompt format. The model receives EVM assembly and optional non-unique signature hints, and must return compiler settings, self-contained Solidity code, and the target contract name in fixed blocks.}
\label{fig:prompt}
\end{figure}
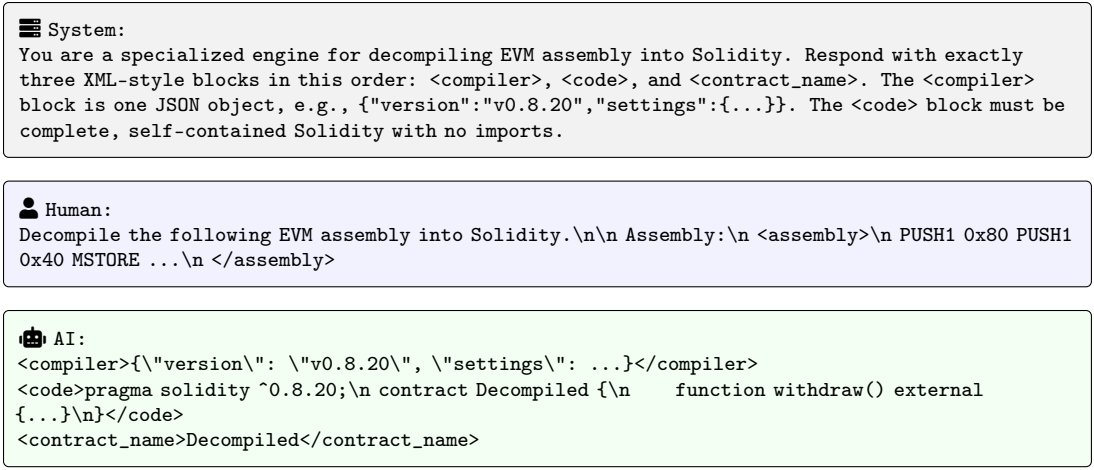

Figure~\ref{fig:prompt} shows the prompt format used for evaluation, and Appendix~\ref{app:prompt-decompilation} gives the complete zero-shot decompilation prompt. The prompt asks the model to return Solidity code, compiler settings, and a target contract name. Together, these fields define an unambiguous compilation target: the code must contain all definitions needed for compilation, the compiler block gives the specified compiler version and configurations, and the contract name identifies which compiled artifact should correspond to the input bytecode, since Solidity compilation may produce multiple artifacts from contracts, libraries, interfaces, or dependencies.

The median assembly length is~$39{,}012$ characters, and the largest complete user prompt, including instructions and signature hints, contains~$136{,}311$ characters. Output-token limits vary by model. We do not exclude contracts whose generation stops at the output-token limit. Appendix~\ref{app:cr-context} reports the input-length distributions, output limits, and sensitivity to length termination.

\textbf{Zero-shot compilation repair.}
During evaluation, we observe that some compilation failures from the initial zero-shot decompilation are minor issues, such as invalid literals, duplicate declarations, or small type inconsistencies. We therefore add a zero-shot compilation-repair variant at the compilability stage. For each model, if the initial output has the required format but fails compilation, we invoke the same model once with the generated Solidity code and compiler error message. The complete repair prompt is shown in Appendix~\ref{app:prompt-repair}. The repaired setting evaluates the final artifact after this single repair opportunity, using the repaired output when available and otherwise retaining the original output. We denote variants with repair enabled using the superscript $^\dagger$. This design tests whether a decompiler can use compiler feedback without changing the model family or adding task-specific training.

\textbf{Model selection.}
We select three of the best-performing models on SWE-bench~\cite{swebench}: Claude Opus 4.7~\cite{anthropic2026claudeopus47} and GPT-5.3-Codex~\cite{openai2026gpt53codex} as closed-source frontier baselines, and GLM-5~\cite{glm5} as the open-weight baseline. The two frontier models are evaluated with high reasoning effort. GLM-5 is evaluated in both instruct (non-thinking) and high-thinking modes to assess the impact of extended reasoning on smart contract decompilation. Throughout the paper, Opus 4.7, GPT-5.3-Codex, and GLM-5 refer to the high-thinking configurations, while GLM-5 (instruct) denotes the non-thinking variant. All models are accessed through OpenRouter.\footnote{\url{https://openrouter.ai/}} Reported costs reflect OpenRouter pricing at the time of evaluation and are subject to change. Detailed token usage and cost are provided in Appendix~\ref{app:pricing}.

\textbf{Evaluation tools.}
For compilation and semantic replay, we use \texttt{solc}\footnote{\url{https://docs.soliditylang.org/en/latest/installing-solidity.html}} and \texttt{foundry}.\footnote{\url{https://github.com/foundry-rs/foundry}} Semantic replay requires access to an Ethereum archive node. Archive-node hardware and storage requirements depend on the node implementation and network state, and we refer readers to Reth's system requirements as one concrete reference.\footnote{\url{https://reth.rs/run/system-requirements/}}

\subsection{Benchmark Results}
\label{sec:benchmark-results}
Overall, the closed frontier models outperform the open-weight baseline across the benchmark. GPT-5.3-Codex$^\dagger$ is the strongest model on the structural stages, achieving the best compilation rate and ABI F1. The end-to-end semantic consistency results are more mixed: Opus 4.7$^\dagger$ obtains the best total semantic consistency rate, while GPT-5.3-Codex$^\dagger$ performs best on hard contracts, showing that the best model for producing compilable source and interfaces is not necessarily the best model for preserving semantics. Reasoning helps after the format stage: GLM-5 with high thinking substantially outperforms GLM-5 (instruct) on compilation, ABI recovery, and semantic consistency. In the following, we report detailed results at each stage.

\begin{table}[t]
\centering
\caption{Format completeness by complexity. A contract passes if the response contains parseable compiler settings, self-contained code, and a target contract name. Each complexity stratum contains $200$ contracts. Highlighted cells indicate the best result in the split.}
\begin{tabular}{lcccc}
\toprule
Model & Easy & Medium & Hard & Total \\
\midrule
Opus 4.7 & $167$ ($83.5\%$) & $143$ ($71.5\%$) & $125$ ($62.5\%$) & $435$ ($72.5\%$) \\
GPT-5.3-Codex & \cellcolor{green!20}$196$ ($98.0\%$) & \cellcolor{green!20}$186$ ($93.0\%$) & \cellcolor{green!20}$182$ ($91.0\%$) & \cellcolor{green!20}$564$ ($94.0\%$) \\
GLM-5 & $169$ ($84.5\%$) & $177$ ($88.5\%$) & $174$ ($87.0\%$) & $520$ ($86.7\%$) \\
GLM-5 (instruct) & $180$ ($90.0\%$) & $185$ ($92.5\%$) & $175$ ($87.5\%$) & $540$ ($90.0\%$) \\
\bottomrule
\end{tabular}
\label{tab:format-results}
\end{table}

\textbf{Format completeness.}
Table~\ref{tab:format-results} shows that GPT-5.3-Codex is the most reliable model at following the required output protocol, producing complete outputs for~$564/600$ contracts ($94.0\%$). GLM-5 (instruct) is slightly better than GLM-5 on this purely format-oriented stage, suggesting that extended reasoning does not necessarily improve instruction-format compliance. Opus 4.7 is more likely to return an output that cannot be automatically parsed. Format failures mostly arise from instruction-following issues, such as malformed XML-style blocks, extra text, or placeholder code, rather than from Solidity compilation errors. The drop from easy to hard is most visible for Opus 4.7, whose format rate falls from~$83.5\%$ to~$62.5\%$.

\begin{table}[t]
\centering
\caption{Compilation success by complexity. A contract passes if the recovered Solidity compiles to a valid target artifact. Each complexity stratum contains $200$ contracts. Blue highlights the best unrepaired result, and green highlights the best result after zero-shot repair.}
\begin{tabular}{lcccc}
\toprule
Model & Easy & Medium & Hard & Total \\
\midrule
Opus 4.7 & $149$ ($74.5\%$) & $113$ ($56.5\%$) & $79$ ($39.5\%$) & $341$ ($56.8\%$) \\
Opus 4.7$^\dagger$ & $164$ ($82.0\%$) & $140$ ($70.0\%$) & $111$ ($55.5\%$) & $415$ ($69.2\%$) \\
GPT-5.3-Codex & \cellcolor{blue!12}$162$ ($81.0\%$) & \cellcolor{blue!12}$142$ ($71.0\%$) & \cellcolor{blue!12}$117$ ($58.5\%$) & \cellcolor{blue!12}$421$ ($70.2\%$) \\
GPT-5.3-Codex$^\dagger$ & \cellcolor{green!20}$193$ ($96.5\%$) & \cellcolor{green!20}$178$ ($89.0\%$) & \cellcolor{green!20}$171$ ($85.5\%$) & \cellcolor{green!20}$542$ ($90.3\%$) \\
GLM-5 & $110$ ($55.0\%$) & $67$ ($33.5\%$) & $44$ ($22.0\%$) & $221$ ($36.8\%$) \\
GLM-5$^\dagger$ & $152$ ($76.0\%$) & $137$ ($68.5\%$) & $114$ ($57.0\%$) & $403$ ($67.2\%$) \\
GLM-5 (instruct) & $74$ ($37.0\%$) & $50$ ($25.0\%$) & $20$ ($10.0\%$) & $144$ ($24.0\%$) \\
GLM-5 (instruct)$^\dagger$ & $153$ ($76.5\%$) & $132$ ($66.0\%$) & $101$ ($50.5\%$) & $386$ ($64.3\%$) \\
\bottomrule
\end{tabular}
\label{tab:compile-results}
\end{table}

\textbf{Compilability.}
Table~\ref{tab:compile-results} shows that compilation is the first stage where model differences become large. In the initial zero-shot decompilation setting, GPT-5.3-Codex compiles on~$421/600$ contracts ($70.2\%$), followed by Opus 4.7 at~$341/600$ ($56.8\%$). GLM-5 and GLM-5 (instruct) are much weaker, compiling~$36.8\%$ and~$24.0\%$ of contracts. The most common compilation failures are localized Solidity issues, including invalid address literals, duplicate declarations, type or mutability errors, unresolved identifiers, and stack-too-deep failures. Appendix~\ref{app:compilation-failures} provides the full failure breakdown. As discussed in Section~\ref{sec:evaluation-setup}, we also evaluate a one-step zero-shot repair variant for outputs that have the required format but fail compilation. Repair improves every model. GPT-5.3-Codex$^\dagger$ remains the strongest, reaching~$542/600$ compilable contracts ($90.3\%$). Opus 4.7$^\dagger$ improves to~$69.2\%$, while GLM-5$^\dagger$ and GLM-5 (instruct)$^\dagger$ improve by more than~$30$ percentage points. These gains suggest that many failures are not complete decompilation failures but repairable compiler-level inconsistencies. Nevertheless, the repaired frontier models still outperform the repaired open-weight baselines, indicating that repair narrows but does not close the model-quality gap.

\begin{table}[t]
\centering
\caption{ABI recovery by complexity, reported as micro precision/recall/F1. Blue marks the best unrepaired F1 in each split, and green marks the best repaired F1.}
\resizebox{\linewidth}{!}{%
\begin{tabular}{lcccc}
\toprule
Model & Easy & Medium & Hard & Total \\
\midrule
Opus 4.7 & $0.931/0.707/0.803$ & $0.950/0.529/0.680$ & $0.898/0.347/0.500$ & $0.924/0.461/0.615$ \\
Opus 4.7$^\dagger$ & $0.930/0.752/0.832$ & $0.940/0.658/0.774$ & $0.903/0.506/0.649$ & $0.921/0.593/0.721$ \\
GPT-5.3-Codex & $0.944/0.803/$\colorbox{blue!12}{$0.868$} & $0.973/0.713/$\colorbox{blue!12}{$0.823$} & $0.943/0.546/$\colorbox{blue!12}{$0.691$} & $0.954/0.639/$\colorbox{blue!12}{$0.766$} \\
GPT-5.3-Codex$^\dagger$ & $0.950/0.935/$\colorbox{green!20}{$0.942$} & $0.971/0.889/$\colorbox{green!20}{$0.928$} & $0.928/0.802/$\colorbox{green!20}{$0.861$} & $0.946/0.851/$\colorbox{green!20}{$0.896$} \\
GLM-5 & $0.924/0.535/0.677$ & $0.940/0.332/0.491$ & $0.894/0.172/0.289$ & $0.920/0.279/0.428$ \\
GLM-5$^\dagger$ & $0.910/0.710/0.798$ & $0.909/0.651/0.759$ & $0.846/0.493/0.623$ & $0.880/0.578/0.697$ \\
GLM-5 (instruct) & $0.914/0.361/0.517$ & $0.731/0.217/0.335$ & $0.871/0.093/0.168$ & $0.819/0.174/0.287$ \\
GLM-5 (instruct)$^\dagger$ & $0.880/0.722/0.793$ & $0.824/0.615/0.704$ & $0.844/0.440/0.578$ & $0.844/0.540/0.658$ \\
\bottomrule
\end{tabular}
}
\label{tab:abi-results}
\end{table}

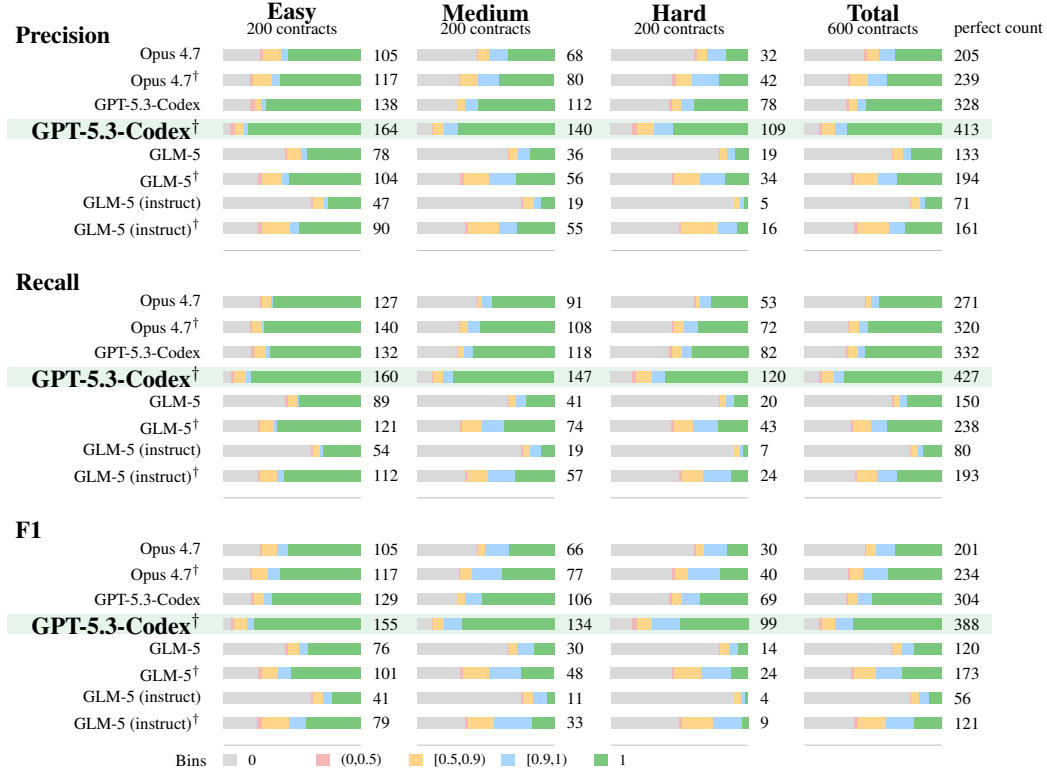
\begin{figure}[t]
\centering
\resizebox{\linewidth}{!}{%
\input{figures/abi_metric_distribution_by_difficulty.tex}
}
\caption{Per-contract ABI recovery distributions by complexity. Each stacked bar groups contracts by precision, recall, or F1 bin. The number at the right of each bar is the count of contracts with perfect score in that metric. The highlighted model has the best ABI F1.}
\label{fig:abi-distribution}
\end{figure}

\textbf{ABI recovery.}
Table~\ref{tab:abi-results} reports ABI recovery by complexity. GPT-5.3-Codex$^\dagger$ achieves the highest F1 in every split, with total micro-F1~$0.896$, while Opus 4.7$^\dagger$ and GLM-5$^\dagger$ also improve substantially after repair. Across models, precision is generally higher than recall, indicating that models are more likely to miss ground-truth functions than to hallucinate many extra callable functions. Repair consistently improves recall because more outputs compile and expose interfaces for comparison, but precision can decrease when the repaired code also introduces extra public functions. Figure~\ref{fig:abi-distribution} shows the per-contract distribution behind these averages. GPT-5.3-Codex$^\dagger$ has the largest mass of high-F1 and perfect-recovery contracts, while all models shift toward lower F1 as complexity increases. The remaining gap suggests that ABI recovery is limited both by compilation success and by the model's ability to infer missing public entry points from bytecode.

\begin{table}[t]
\centering
\caption{Semantic consistency by complexity. Denominators are $2{,}226$ easy, $4{,}774$ medium, $7{,}553$ hard, and $14{,}553$ total public functions.}
\begin{tabular}{lcccc}
\toprule
Model & Easy & Medium & Hard & Total \\
\midrule
Opus 4.7 & \cellcolor{blue!12}$865$ ($38.9\%$) & \cellcolor{blue!12}$1158$ ($24.3\%$) & \cellcolor{blue!12}$1248$ ($16.5\%$) & \cellcolor{blue!12}$3271$ ($22.5\%$) \\
Opus 4.7$^\dagger$ & \cellcolor{green!20}$936$ ($42.0\%$) & \cellcolor{green!20}$1521$ ($31.9\%$) & $1827$ ($24.2\%$) & \cellcolor{green!20}$4284$ ($29.4\%$) \\
GPT-5.3-Codex & $732$ ($32.9\%$) & $943$ ($19.8\%$) & $1176$ ($15.6\%$) & $2851$ ($19.6\%$) \\
GPT-5.3-Codex$^\dagger$ & $842$ ($37.8\%$) & $1334$ ($27.9\%$) & \cellcolor{green!20}$1902$ ($25.2\%$) & $4078$ ($28.0\%$) \\
GLM-5 & $533$ ($23.9\%$) & $565$ ($11.8\%$) & $344$ ($4.6\%$) & $1442$ ($9.9\%$) \\
GLM-5$^\dagger$ & $740$ ($33.2\%$) & $1170$ ($24.5\%$) & $1105$ ($14.6\%$) & $3015$ ($20.7\%$) \\
GLM-5 (instruct) & $340$ ($15.3\%$) & $317$ ($6.6\%$) & $183$ ($2.4\%$) & $840$ ($5.8\%$) \\
GLM-5 (instruct)$^\dagger$ & $654$ ($29.4\%$) & $903$ ($18.9\%$) & $1013$ ($13.4\%$) & $2570$ ($17.7\%$) \\
\bottomrule
\end{tabular}
\label{tab:semantic-results}
\end{table}

\begin{table}[t]
\centering
\caption{Perfect semantic consistency by complexity. A contract is counted as perfect only if all benchmarked public functions are recovered, all replayed semantic checkpoints match, and no spurious function is recovered.}
\begin{tabular}{lcccc}
\toprule
Model & Easy & Medium & Hard & Total \\
\midrule
Opus 4.7 & $25$ ($12.5\%$) & \cellcolor{blue!12}$6$ ($3.0\%$) & \cellcolor{blue!12}$2$ ($1.0\%$) & \cellcolor{blue!12}$33$ ($5.5\%$) \\
Opus 4.7$^\dagger$ & $33$ ($16.5\%$) & \cellcolor{green!20}$7$ ($3.5\%$) & \cellcolor{green!20}$2$ ($1.0\%$) & \cellcolor{green!20}$42$ ($7.0\%$) \\
GPT-5.3-Codex & \cellcolor{blue!12}$30$ ($15.0\%$) & $3$ ($1.5\%$) & $0$ ($0.0\%$) & \cellcolor{blue!12}$33$ ($5.5\%$) \\
GPT-5.3-Codex$^\dagger$ & \cellcolor{green!20}$37$ ($18.5\%$) & $4$ ($2.0\%$) & $0$ ($0.0\%$) & $41$ ($6.8\%$) \\
GLM-5 & $14$ ($7.0\%$) & $0$ ($0.0\%$) & $0$ ($0.0\%$) & $14$ ($2.3\%$) \\
GLM-5$^\dagger$ & $17$ ($8.5\%$) & $1$ ($0.5\%$) & $0$ ($0.0\%$) & $18$ ($3.0\%$) \\
GLM-5 (instruct) & $4$ ($2.0\%$) & $0$ ($0.0\%$) & $0$ ($0.0\%$) & $4$ ($0.7\%$) \\
GLM-5 (instruct)$^\dagger$ & $6$ ($3.0\%$) & $0$ ($0.0\%$) & $0$ ($0.0\%$) & $6$ ($1.0\%$) \\
\bottomrule
\end{tabular}
\label{tab:perfect-semantic-results}
\end{table}

\textbf{Semantic consistency.}
Table~\ref{tab:semantic-results} shows the largest gap between interface recovery and semantic consistency. Opus 4.7$^\dagger$ achieves the highest total semantic consistency rate, with~$4{,}284/14{,}553$ public functions consistent ($29.4\%$). GPT-5.3-Codex$^\dagger$ is close overall ($28.0\%$) and performs best on the hard split ($25.2\%$), but its stronger compilation and ABI recovery do not translate into the best overall semantic consistency score. The open-weight models remain substantially behind the frontier models, although GLM-5 with reasoning improves over GLM-5 (instruct). Semantic consistency also drops sharply with complexity, reflecting the challenge of preserving state updates, access-control logic, revert conditions, and event emissions in larger contracts. Table~\ref{tab:perfect-semantic-results} gives the stricter contract-level view. Even the best setting perfectly recovers only~$42/600$ contracts, and almost no hard contracts are recovered perfectly. These results show why SCDBench reports semantic consistency as a separate final stage: high compilation and ABI recovery are necessary but far from sufficient for faithful decompilation.

Traditional decompilers generally do not produce self-contained, compilable Solidity, so they do not fit the full SCDBench pipeline. We nevertheless run two representative tools on our dataset and report scope-limited ABI recovery results in Appendix~\ref{app:traditional-decompilers}.

\subsection{Validation and Sensitivity}
\label{sec:cr-validation}
\textbf{Empirical validation of the complexity strata.}
To understand whether the complexity strata reflect observed decompilation difficulty, we examine the relationship between the complexity score and model performance. Across the four repaired model settings, the complexity score has Spearman correlations of~$-0.358$, $-0.377$, and~$-0.327$ with average per-contract compilation success, ABI F1, and semantic consistency, respectively. All five score components are negatively correlated with these three outcomes. The negative associations for the combined score persist after accounting for model, prompt length, and public-function count. The relationship with format completeness is weaker. We suspect this is because current models are capable of following output instructions and generating compilable Solidity, while the generated code does not necessarily recover the original bytecode's logic. These results provide empirical support for using the score to construct complexity strata that reflect observed decompilation difficulty. Appendix~\ref{app:cr-complexity} reports the component-level correlations and adjusted analysis.

\textbf{Sensitivity to signature candidates.}
To understand how much decompilation performance depends on public signature hints, we sample~$60$ contracts from the original benchmark, with~$20$ from each complexity stratum, and evaluate GPT-5.3-Codex without the external function-selector and event-signature mappings. Assembly, intrinsic selectors and event topics, model settings, and checkpoints were unchanged. After repair, ABI F1 changes from~$0.827$ with hints to~$0.430$ without hints, and semantic consistency from~$31.2\%$ to~$15.5\%$, although compilation rises from~$85.0\%$ to~$91.7\%$. These results suggest that public signature hints help semantic recovery. The higher compilation rate without hints illustrates that code can compile successfully while failing to reproduce the original contract's behavior. We report the detailed results in Appendix~\ref{app:cr-signature-hints}.

\textbf{Checkpoint effectiveness.}
We further analyze how effectively the semantic checkpoints detect behavioral mismatches. From the decompilation outputs that reach replay under the four repaired model settings, we find that~$344{,}055$ of~$564{,}261$ semantic replay checks ($61.0\%$) detect a behavioral mismatch. At the function level, $22{,}837$ of~$36{,}784$ decompiled functions tested ($62.1\%$) fail at least one checkpoint. We detail the effectiveness of non-random cases, storage checks, and event-log checks in Appendix~\ref{app:cr-test-effectiveness}.

\textbf{Coverage sensitivity.}
To assess whether low-coverage contracts affect our conclusions, we recompute semantic consistency on the~$239$ contracts with at least~$80\%$ code coverage. Repaired Opus 4.7, GPT-5.3-Codex, GLM-5, and GLM-5 (instruct) achieve semantic consistency of~$29.7\%$, $29.5\%$, $21.4\%$, and~$17.2\%$, respectively. Our main findings remain consistent on this high-coverage subset: frontier models achieve higher semantic consistency than open-weight models, and GLM-5 with reasoning outperforms its instruct variant. We present detailed results for subsets defined by different code-coverage thresholds in Appendix~\ref{app:cr-coverage-sensitivity}.

\textbf{A second 600-contract sample.}
To test stability across a different sample, we reran all four model settings, including one-step repair, on a second~$600$-contract sample with the same~$200/200/200$ complexity strata. It shares no contracts or selected near-duplicate clusters with the original sample. We generated~$226{,}138$ replayable semantic cases over~$14{,}578$ callable entries. The second sample reproduces the main structural and semantic findings. Repaired GPT-5.3-Codex again leads compilation and ABI recovery ($93.3\%$ and~$0.917$ F1), while repaired Opus 4.7 has the highest semantic consistency ($32.0\%$, compared with GPT-5.3-Codex's~$30.0\%$). One-step repair improves every model, but full recovery remains rare: Opus 4.7 and GPT-5.3-Codex each recover~$36/600$ contracts. Semantic consistency decreases from easy to hard for every model. Appendix~\ref{app:cr-second-sample} reports all five result tables, including initial and repaired settings. These results support the stability of our main findings across the two contract samples.

\section{Conclusion and Discussion}
We present a systematic benchmark for smart contract decompilation, combining a curated dataset of~$600$ real-world contracts with a staged evaluation framework. The dataset balances real-world grounding with controlled diversity, spans multiple complexity strata, and supports reproducible semantic consistency testing through function-level semantic checkpoints. The four metrics, format completeness, compilability, ABI recovery accuracy, and semantic consistency, form a progressive pipeline that reveals both syntactic and semantic strengths and weaknesses.

Our results provide three main insights. First, smart contract decompilation benefits from the development of reasoning capacities in frontier models. The closed-source frontier models substantially outperform the open-weight baseline, and the high-thinking GLM-5 configuration consistently outperforms its non-thinking variant after the format stage on the original sample. This suggests that decompilation benefits from extended reasoning over bytecode structure, compiler constraints, and interface hints. Second, zero-shot compilation repair significantly improves performance while adding only a modest amount of cost, as shown by the token and cost analysis in Appendix~\ref{app:pricing}. This indicates that many failures are repairable compiler-level inconsistencies rather than complete decompilation failures, and points to the potential of agentic decompiler designs that decompose decompilation into subtasks such as initial source recovery, compiler-guided repair, ABI checking, and semantic replay. Third, structural success is not semantic success, and perfect semantic consistency remains an open problem. The model with the strongest compilation and ABI recovery is not necessarily the model with the strongest semantic consistency, and even the best repaired model perfectly recovers only a small fraction of contracts, especially in the hard split. The gains from one-step repair motivate tool-augmented decompilation, while the remaining semantic gap highlights the need for feedback beyond compilation.

Overall, SCDBench establishes a common ground for reproducible evaluation and highlights where future smart contract decompilers need to improve: robust compilation, accurate interface recovery, and above all semantic consistency.





{
\small
\bibliographystyle{plainnat}
\bibliography{references}
}

\appendix

\section{Dataset Construction and External Validity}
\label{app:dataset-construction}
\subsection{Near-Duplicate Filtering}
\label{app:near-duplicate-filtering}
We use deterministic fingerprints rather than neural embeddings for hard duplicate exclusion. The bytecode fingerprints are computed after stripping Solidity metadata. The strict bytecode-structure hash removes all \texttt{PUSH} operands while preserving opcode identities, so \texttt{PUSH1} and \texttt{PUSH32} remain distinct. The coarse bytecode-structure hash also removes operands, but maps opcode families such as \texttt{PUSHx}, \texttt{DUPx}, \texttt{SWAPx}, and \texttt{LOGx} to shared family symbols. The source-skeleton hash tokenizes Solidity source while replacing identifiers and literals with placeholders.

Near-duplicate clusters are formed by unioning exact fingerprint collisions. The selected profile always merges exact strict bytecode-structure matches, and also merges exact coarse-bytecode or source-skeleton matches when the contracts fall within the same~$10\%$ logarithmic bytecode-length band. The length guard prevents a shared template skeleton from merging contracts with materially different compiled scale. This selected profile yields~$333{,}062$ near-duplicate clusters from the~$772{,}736$ exact-bytecode-deduplicated contracts, and no two benchmark contracts are allowed to share a cluster.

\subsection{Complexity Scoring}
\label{app:difficulty-scoring}
We assign complexity scores to one deterministic representative from each near-duplicate cluster. Continuous features are transformed by \texttt{log1p} and robustly scaled between their~$5$th and~$95$th percentiles over the cluster-representative universe:
\[
z_m(x)=\mathrm{clip}\left(\frac{\log(1+x)-p05_m}{p95_m-p05_m},0,1\right).
\]
The final score ranges from~$0$ to~$100$:
\[
\mathrm{score}=100\cdot(0.30B+0.20C+0.20I+0.15S+0.15L),
\]
where $B$ is bytecode scale, $C$ is control flow, $I$ is interface and source structure, $S$ is state interaction, and $L$ is low-level Solidity features. Table~\ref{tab:difficulty-components} lists the feature groups. Source-derived features are used only for benchmark construction and are not provided as model input during decompilation.

\begin{table}[h]
\centering
\caption{Complexity-score components used for benchmark stratification.}
\begin{tabular}{@{}ll@{}}
\toprule
Component & Features \\
\midrule
Bytecode scale & bytecode bytes, opcode count \\
Control flow & cyclomatic proxy, block-count proxy, control-flow opcode count \\
Interface/source & signature count, declaration count, nonblank source lines \\
State interaction & storage ops, external calls, create ops, log ops \\
Low-level Solidity features & inline assembly, delegatecall, fallback, receive, create2, selfdestruct \\
\bottomrule
\end{tabular}
\label{tab:difficulty-components}
\end{table}

\subsection{Sampling}
\label{app:sampling}
The cluster-representative universe is split into three equal-sized complexity strata by score tertiles, producing~$111{,}021$ easy candidates,~$111{,}020$ medium candidates, and~$111{,}021$ hard candidates. Before sampling, candidates with no nontrivial bytecode or source signal are excluded, and selected contracts with empty or no-op-like externally callable functions are rejected and resampled.

Within each complexity stratum, candidates are stratified by compiler minor version and coarse contract label. The~$200$ contracts per bin are allocated across strata using largest-remainder rounding with square-root stratum weights, $w_s=\sqrt{n_s}$. This compresses dominant strata while avoiding over-allocation to rare strata. Within each stratum, selected contracts are spread across the local complexity-score range using quantile positions with deterministic hash tie-breaks.

\subsection{Sample Characteristics}
\label{app:sample-characteristics}
The final sample covers all five Solidity compiler minor versions from~v0.4 to~v0.8 and spans a broad range of code characteristics. Table~\ref{tab:sample-diversity} shows that public-function counts, bytecode size, source length, control-flow complexity, and storage operations increase monotonically across the complexity strata. This indicates that the benchmark contains both small contracts and substantially larger stateful contracts rather than a narrow slice of the corpus.

\begin{table}[h]
\centering
\caption{Code-characteristic diversity indicators for the~$600$-contract benchmark sample. Complexity columns report medians within each split.}
\resizebox{\linewidth}{!}{%
\begin{tabular}{lrrrrrrr}
\toprule
Split & Contracts & Public functions & Compiler minors & Bytecode bytes & Source lines & CFG proxy & Storage ops \\
\midrule
Easy & $200$ & $2{,}226$ & $5$ & $2{,}924$ & $152$ & $49.5$ & $20$ \\
Medium & $200$ & $4{,}774$ & $5$ & $7{,}699$ & $419$ & $132$ & $72.5$ \\
Hard & $200$ & $7{,}553$ & $5$ & $15{,}342$ & $918$ & $252$ & $120.5$ \\
Total & $600$ & $14{,}553$ & $5$ & $7{,}782$ & $441$ & $130.5$ & $60$ \\
\bottomrule
\end{tabular}
}
\label{tab:sample-diversity}
\end{table}

\subsection{External Validity}
\label{app:cr-external-validity}

To assess the scope of verified-source sampling, we randomly sampled
1,000,000 deployed Ethereum contracts with non-empty bytecode. After
removing exact bytecode duplicates, we retain 37,653 distinct contract bytecodes.
We report statistics both before deduplication, counting every sampled
contract, and after deduplication, counting each distinct bytecode once.
We compare the deduplicated sample with the 772,736 distinct bytecodes
in the verified corpus. The deployed sample was drawn without
verification labels and is not a labeled unverified-contract accuracy set.

\begin{table}[htbp]

\centering
\normalsize
\setlength{\tabcolsep}{3pt}
\caption{Bytecode statistics for deployed contracts,
the verified corpus, and SCDBench. Opcode rows report the fraction of bytecodes
containing the indicated operation.}
\label{tab:cr-external-validity}
\begin{tabular}{@{}p{.33\linewidth}rrrr@{}}
\toprule
Statistic & Deployed & Unique deployed & Unique verified & SCDBench \\
\midrule
Median bytecode bytes & 45 & 2,100 & 6,831 & 7,830 \\
Canonical EIP-1167 minimal proxy & 44.63\% & 16.94\% & 0.00\% & 0.00\% \\
Contains SLOAD (storage read) & 19.54\% & 64.45\% & 96.41\% & 93.50\% \\
Contains SSTORE (storage write) & 8.44\% & 54.85\% & 87.52\% & 89.83\% \\
Contains DELEGATECALL & 56.39\% & 36.96\% & 11.40\% & 9.83\% \\
\bottomrule
\end{tabular}
\end{table}

Minimal proxies are tiny contracts with minimal logic beyond delegating
execution to another contract. Repeated deployments of these proxies explain
much of the difference before deduplication.
After exact bytecode duplicates are removed, deployed contracts remain smaller
than verified contracts, with median sizes of 2,100 versus 6,831 bytes. They
use DELEGATECALL more often and storage operations less often. The aggregate
opcode distributions nevertheless have a total-variation distance of 0.035,
where zero means identical distributions.

Compiler versions can be recovered from bytecode metadata for 54.56\% of the
distinct deployed bytecodes. Among those with recovered versions, 75.46\% use
Solidity 0.8, compared with 77.45\% of the verified corpus. Metadata recovery
depends on its presence and format, so these results do not establish that
the complete compiler distributions are identical.

Overall, the two populations have similar aggregate opcode distributions
but differ in contract size and proxy prevalence. SCDBench therefore targets
diversity within source-verified Solidity contracts rather than representing
the distribution of all deployed contracts.

\FloatBarrier
\section{Semantic Test Construction and Replay}
\label{app:test-cases}
For each benchmark contract, we compile the ground-truth Solidity source to obtain its ABI and enumerate public and external functions, fallback entries, and receive entries. Test cases are stored as a map from function signature to a list of concrete cases. This organization supports the staged evaluation pipeline: semantic replay is applied to each function whose signature is recovered by the decompiler, even when other functions in the same contract are missing.

The input generator combines four sources: deterministic random values for supported ABI types, static hints from constants and literals in the ground-truth source and bytecode, boundary values such as zero, one, empty byte strings, and threshold-like integers, and short stateful setup sequences for functions that require prior state. Functions with inputs or payable behavior receive ten deterministic ABI-valid cases and other functions receive one, supplemented by up to four boundary, sixteen static-hint, and two stateful cases. Static hints are used only for test generation. Duplicate calldata/value pairs are excluded when adding boundary and static-hint cases. Payable inputs include zero, one wei, and bounded nonzero values.

Each case starts from the same freshly deployed target state on a fixed mainnet fork, restored by snapshot rollback. A stateful case uses one setup call followed by the measured call, both from the same funded harness-contract sender. These setup calls target common smart contract patterns such as owner checks, role checks, allowance updates, balance updates, and write-then-read workflows. External calls execute against the fork. Each checkpoint records the concrete input and any setup calls, together with the measured call's success/revert status, exact return/revert bytes, ordered logs, and target-contract storage changes. Semantic replay executes the same case against the decompiler output and compares these observables with the recorded checkpoint, with storage changes checked for state-changing functions. Reverts are kept as valid execution outcomes.

The comparison excludes external-contract storage, balances, and gas. There is no special proxy resolution, historical target-storage reconstruction, or failure minimization. Tests do not cover longer call histories, multiple callers, or changing block context.

\FloatBarrier
\section{Comparison with Smart-Contract Fuzzing}
\label{app:cr-fuzzing}

Smart-contract fuzzers and SCDBench share techniques for input generation and
stateful execution, but their objectives and oracles differ. Fuzzers generally search
for vulnerabilities by exploring transaction sequences and parameters that
trigger security oracles. SCDBench replays matched calls against original and
decompiled contracts to compare returns, reverts, logs, and storage changes.
A vulnerability finding is therefore not interchangeable with a semantic
mismatch.

\paragraph{Coverage experiment.}
We ran ItyFuzz on a fixed subset of $20$ easy, $20$ medium, and $20$ hard
contracts from the original benchmark.\footnote{We used ItyFuzz commit
\href{https://github.com/fuzzland/ityfuzz/commit/80e3b44170573b592cd53b2c5ee9f36b21032b16}{\texttt{80e3b4417057}}.}
We allocated ItyFuzz $4$ seconds per easy contract, $9$ per medium contract,
and $16$ per hard contract. These budgets were set using the measured time
to replay SCDBench's semantic checkpoints in each complexity stratum as a reference.
Table~\ref{tab:cr-fuzzer-coverage} compares code coverage on the selected contracts.

\begin{table}[htbp]
\centering

\normalsize
\caption{Code coverage for selected contracts with measurements available for both methods.
ItyFuzz budgets are $4$, $9$, and $16$ seconds for easy, medium, and hard contracts.}
\label{tab:cr-fuzzer-coverage}
\begin{tabular}{@{}lrrrr@{}}
\toprule
Method & Easy & Medium & Hard & Overall \\
\midrule
SCDBench checkpoints & $88.8\%$ & $45.7\%$ & $72.1\%$ & $67.9\%$ \\
ItyFuzz & $30.2\%$ & $25.7\%$ & $20.6\%$ & $24.0\%$ \\
\bottomrule
\end{tabular}
\end{table}

\paragraph{Timing and execution scope.}
It is important to note that SCDBench timings measure execution of already
generated checkpoints, excluding checkpoint generation, compilation, and
coverage instrumentation. The ItyFuzz
campaign spends its budget searching for inputs and sequences. The table is
therefore a descriptive coverage comparison under these configurations, not
an equal-total-cost comparison of test generators. SCDBench sequences contain
at most two calls, whereas ItyFuzz's exported sequences reached depth $21$.
The ItyFuzz coverage experiment does not use a differential oracle, so it does
not produce a directly comparable semantic-mismatch count.

\paragraph{Design differences.}
First, default ItyFuzz focuses on state-changing functions and does not add
view functions to its fuzzing corpus. The three view-only contracts in the
selected subset consequently receive zero ItyFuzz coverage. Second, its
parameter mutations and sequence exploration seek vulnerable states rather
than systematically providing matched differential inputs for each public
function. Third, ItyFuzz caches intermediate states so that later mutations
can resume from promising states. SCDBench uses at most one setup call before
the measured target call. These differences explain why the comparison does
not establish superiority for either method's intended task.

\FloatBarrier
\section{Detailed Evaluation Results}
\label{app:evaluation-diagnostics}
\subsection{Conditional Evaluation Diagnostics}
\label{app:cr-diagnostics}

We report cumulative scores in Section~\ref{sec:benchmark-results}, which
measure end-to-end decompilation performance across the full benchmark.
Table~\ref{tab:cr-conditional-diagnostics} reports the corresponding conditional metrics.

\begin{table}[htbp]
\centering

\normalsize
\setlength{\tabcolsep}{4pt}
\caption{Conditional metrics before and after one-step repair.
$\dagger$ denotes repair. Compilation and semantic columns include their denominators.}
\label{tab:cr-conditional-diagnostics}
\begin{tabular}{@{}lccc@{}}
\toprule
Model & Compilation (valid format) &
ABI F1 (compiled) & Semantics (replayed entries) \\
\midrule
Opus 4.7 & $341/435$ ($78.4\%$) & $.946$ & $3{,}271/6{,}594$ ($49.6\%$) \\
Opus 4.7$^\dagger$ & $415/435$ ($95.4\%$) & $.939$ & $4{,}284/8{,}498$ ($50.4\%$) \\
GPT-5.3-Codex & $421/564$ ($74.6\%$) & $.950$ & $2{,}851/9{,}187$ ($31.0\%$) \\
GPT-5.3-Codex$^\dagger$ & $542/564$ ($96.1\%$) & $.946$ & $4{,}078/12{,}227$ ($33.4\%$) \\
GLM-5 & $221/520$ ($42.5\%$) & $.917$ & $1{,}442/4{,}013$ ($35.9\%$) \\
GLM-5$^\dagger$ & $403/520$ ($77.5\%$) & $.892$ & $3{,}015/8{,}302$ ($36.3\%$) \\
GLM-5 (instruct) & $144/540$ ($26.7\%$) & $.855$ & $840/2{,}497$ ($33.6\%$) \\
GLM-5 (instruct)$^\dagger$ & $386/540$ ($71.5\%$) & $.871$ & $2{,}570/7{,}757$ ($33.1\%$) \\
\bottomrule
\end{tabular}
\end{table}

We find that strong ABI recovery does not necessarily translate into correct contract
behavior. After repair, conditional ABI F1 ranges from $.871$ to $.946$,
yet only $33.1\%$--$50.4\%$ of replayed callable entries are semantically
consistent. Repair substantially increases the number of replayable entries,
while conditional semantic consistency changes little. For example,
GPT-5.3-Codex's replayed entries increase from $9{,}187$ to $12{,}227$,
whereas its conditional semantic consistency rises from $31.0\%$ to $33.4\%$.

\subsection{Compilation Failure Analysis}
\label{app:compilation-failures}
Table~\ref{tab:compilation-failure-reasons} reports the primary compiler failure reason for each before-repair output that passed format validation but failed Solidity compilation. The dominant failures are localized Solidity issues such as invalid address literals, duplicate declarations, type or mutability errors, unresolved identifiers, and parser errors. Many of these failures are fixed by a single same-model repair call.

\begin{table}[h]
\centering
\scriptsize
\caption{Compilation failure reasons before repair and failures fixed by zero-shot repair. Each cell reports failures before repair, with fixed failures in parentheses. Model rows omitted within a failure reason have zero failures in that class.}
\label{tab:compilation-failure-reasons}
\begin{tabular}{p{0.38\linewidth}p{0.17\linewidth}rrrr}
\toprule
Failure reason & Model & Easy & Medium & Hard & Total \\
\midrule
Invalid address literal/checksum & Opus 4.7 & $11$ ($10$) & $10$ ($9$) & $13$ ($5$) & $34$ ($24$) \\
 & GPT-5.3-Codex & $13$ ($12$) & $24$ ($22$) & $30$ ($24$) & $67$ ($58$) \\
 & GLM-5 & $11$ ($9$) & $22$ ($16$) & $14$ ($9$) & $47$ ($34$) \\
 & GLM-5 (instruct) & $8$ ($7$) & $12$ ($7$) & $14$ ($10$) & $34$ ($24$) \\
\addlinespace
Duplicate declarations/name collisions & Opus 4.7 & $1$ ($0$) & $1$ ($1$) & $7$ ($4$) & $9$ ($5$) \\
 & GPT-5.3-Codex & -- & -- & $3$ ($2$) & $3$ ($2$) \\
 & GLM-5 & $14$ ($7$) & $44$ ($33$) & $35$ ($21$) & $93$ ($61$) \\
 & GLM-5 (instruct) & $41$ ($31$) & $52$ ($30$) & $64$ ($33$) & $157$ ($94$) \\
\addlinespace
Type/mutability errors & Opus 4.7 & $1$ ($1$) & $4$ ($4$) & $10$ ($9$) & $15$ ($14$) \\
 & GPT-5.3-Codex & $2$ ($1$) & $3$ ($2$) & $6$ ($6$) & $11$ ($9$) \\
 & GLM-5 & $14$ ($10$) & $10$ ($8$) & $25$ ($14$) & $49$ ($32$) \\
 & GLM-5 (instruct) & $20$ ($18$) & $26$ ($21$) & $21$ ($14$) & $67$ ($53$) \\
\addlinespace
Unresolved identifiers/members & Opus 4.7 & $1$ ($1$) & $2$ ($2$) & $6$ ($5$) & $9$ ($8$) \\
 & GPT-5.3-Codex & -- & $1$ ($1$) & $4$ ($3$) & $5$ ($4$) \\
 & GLM-5 & $7$ ($6$) & $17$ ($7$) & $37$ ($19$) & $61$ ($32$) \\
 & GLM-5 (instruct) & $19$ ($15$) & $22$ ($13$) & $25$ ($16$) & $66$ ($44$) \\
\addlinespace
Parser errors & Opus 4.7 & $1$ ($1$) & $2$ ($2$) & $2$ ($2$) & $5$ ($5$) \\
 & GPT-5.3-Codex & $3$ ($2$) & -- & $3$ ($2$) & $6$ ($4$) \\
 & GLM-5 & $8$ ($7$) & $12$ ($4$) & $14$ ($4$) & $34$ ($15$) \\
 & GLM-5 (instruct) & $15$ ($6$) & $14$ ($3$) & $26$ ($6$) & $55$ ($15$) \\
\addlinespace
Invalid compiler settings & Opus 4.7 & -- & $2$ ($2$) & -- & $2$ ($2$) \\
 & GPT-5.3-Codex & $13$ ($13$) & $7$ ($4$) & -- & $20$ ($17$) \\
 & GLM-5 & $4$ ($2$) & $2$ ($0$) & $3$ ($1$) & $9$ ($3$) \\
 & GLM-5 (instruct) & $2$ ($1$) & $4$ ($3$) & $2$ ($0$) & $8$ ($4$) \\
\addlinespace
Stack too deep & Opus 4.7 & $2$ ($1$) & $8$ ($6$) & $7$ ($6$) & $17$ ($13$) \\
 & GPT-5.3-Codex & $3$ ($3$) & $8$ ($6$) & $19$ ($17$) & $30$ ($26$) \\
 & GLM-5 & -- & $2$ ($1$) & $2$ ($2$) & $4$ ($3$) \\
 & GLM-5 (instruct) & -- & $4$ ($4$) & $3$ ($2$) & $7$ ($6$) \\
\addlinespace
Inheritance/override/abstractness & Opus 4.7 & $1$ ($1$) & $1$ ($1$) & $1$ ($1$) & $3$ ($3$) \\
\addlinespace
Syntax errors & GPT-5.3-Codex & -- & $1$ ($1$) & -- & $1$ ($1$) \\
 & GLM-5 (instruct) & $1$ ($1$) & -- & -- & $1$ ($1$) \\
\addlinespace
Target artifact/bytecode errors & GLM-5 & $1$ ($1$) & -- & -- & $1$ ($1$) \\
\addlinespace
Other & GLM-5 & -- & $1$ ($1$) & -- & $1$ ($1$) \\
 & GLM-5 (instruct) & -- & $1$ ($1$) & -- & $1$ ($1$) \\
\bottomrule
\end{tabular}
\end{table}

\FloatBarrier
\subsection{Traditional Decompiler ABI Baselines}
\label{app:traditional-decompilers}
Traditional smart contract decompilers remain strong at recovering function entry points from bytecode. This is expected because many four-byte selectors are embedded directly in dispatcher logic and can often be matched precisely against public signature databases. However, these tools usually emit low-level intermediate representations or pseudocode rather than self-contained Solidity. As a result, readability remains a concern, and their outputs cannot generally be compiled or passed through the semantic consistency stages of SCDBench.

We therefore evaluate Gigahorse and Heimdall-rs only as scope-limited ABI recovery baselines, with results shown in Table~\ref{tab:traditional-decompiler-abi}. Selector-level evaluation compares recovered four-byte function selectors against ground-truth selectors. When typed prototypes are available, prototype-level evaluation also requires the recovered argument type sequence to match.

\begin{table}[h]
\centering
\caption{ABI recovery for traditional decompilers. These tools recover function selectors precisely, but they do not generally produce compilable Solidity for the full SCDBench pipeline.}
\begin{tabular}{lccc}
\toprule
Tool & Successful runs & Selector micro F1 & Prototype micro F1 \\
\midrule
Gigahorse & $600/600$ & $0.991$ & n/a \\
Heimdall-rs & $599/600$ & $0.996$ & $0.777$ \\
\bottomrule
\end{tabular}
\label{tab:traditional-decompiler-abi}
\end{table}

Table~\ref{tab:traditional-decompiler-abi} confirms that selector recovery is comparatively mature for traditional decompilers. Gigahorse successfully runs on all~$600$ contracts and reaches selector micro-F1~$0.991$, while Heimdall-rs runs on~$599/600$ contracts and reaches selector micro-F1~$0.996$. Heimdall-rs also recovers typed prototypes with micro-F1~$0.777$, showing that argument-type recovery is harder than recovering four-byte selectors alone. These results indicate that traditional decompilers remain useful for recovering callable interfaces, but selector recovery alone does not address the central goals of source-level decompilation: readable Solidity, successful recompilation, and semantic consistency with the original bytecode.

\FloatBarrier
\section{Prompt Templates}
\label{app:prompts}
This appendix lists the complete prompt templates used in the zero-shot decompilation and zero-shot compilation-repair experiments. Braced placeholders denote benchmark-specific content inserted by the evaluation script.

\subsection{Zero-Shot Decompilation Prompt}
\label{app:prompt-decompilation}
\textbf{System prompt.}
\begin{lstlisting}
You are a specialized engine for decompiling EVM assembly into Solidity.

Your entire response must consist of exactly three XML-style blocks in this exact order:
<compiler></compiler>
<code></code>
<contract_name></contract_name>

The <compiler> block must contain a single JSON object. Example:
<compiler>
{"version":"v0.8.20","settings":{"optimizer":{"enabled":true,"runs":200},"evmVersion":"shanghai"}}
</compiler>

Rules:
- Do not output Markdown fences, explanations, notes, or any text outside the three required blocks.
- The three blocks must be top-level blocks. Do not place <compiler>, <code>, or <contract_name> tags inside another block.
- Close the <code> block before starting the <contract_name> block.
- Put complete, self-contained Solidity source in the <code> block, including the pragma and the main contract.
- Do not use import statements. Do not reference external packages such as OpenZeppelin, Solmate, Uniswap, Chainlink, or project-local files.
- If a library, interface, base contract, or error type is needed, define the minimal required version directly inside the <code> block.
- The <code> block must be compilable as a single Solidity file using only built-in Solidity/EVM features and definitions present in that same block.
- Infer a plausible compiler version and settings from the assembly when possible. If a field cannot be inferred, use null for that field in the compiler JSON.
- Use the function and event signature mappings only as uncertain hints. They may be incomplete, ambiguous, or non-unique.
- Prefer readable, compiling Solidity over literal opcode-by-opcode comments.
\end{lstlisting}

\textbf{User prompt.}
\begin{lstlisting}
Decompile the following EVM assembly into Solidity.

Assembly:
<assembly>
{assembly}
</assembly>

Possible non-unique, uncertain function/event signature mappings:
Functions:
{selector}: {signature_1} | {signature_2} | ...
Events:
{event_hash}: {event_signature_1} | {event_signature_2} | ...
\end{lstlisting}
The signature-mapping block is omitted when no function or event hints are available.

\subsection{Zero-Shot Compilation-Repair Prompt}
\label{app:prompt-repair}
\textbf{System prompt.}
\begin{lstlisting}
You are a specialized Solidity compilation repair engine.

Your entire response must consist of exactly three XML-style blocks in this exact order:
<compiler></compiler>
<code></code>
<contract_name></contract_name>

The <compiler> block must contain a single JSON object. Example:
<compiler>
{"version":"v0.8.20","settings":{"optimizer":{"enabled":true,"runs":200},"evmVersion":"shanghai"}}
</compiler>

Rules:
- Do not output Markdown fences, explanations, notes, or any text outside the three required blocks.
- Repair the provided decompilation so it compiles as a single self-contained Solidity file.
- Do not use import statements or external packages.
- If a library, interface, base contract, or error type is needed, define a minimal version directly in the <code> block.
- Fix compiler JSON mistakes when the compiler error indicates invalid settings.
- Preserve the intended ABI and semantics of the initial decompilation as much as possible.
- The <contract_name> block must name the contract artifact that should be compiled and evaluated.
\end{lstlisting}

\textbf{User prompt.}
\begin{lstlisting}
A previous EVM-to-Solidity decompilation failed to compile. Repair it using the compiler diagnostics below.

Initial decompilation output:
<initial_output>
{initial_decompilation_output}
</initial_output>

Compilation error from solc:
<compilation_error>
{solc_compilation_error}
</compilation_error>

Return only the repaired <compiler>, <code>, and <contract_name> blocks. Do not explain the changes.
\end{lstlisting}

\FloatBarrier
\section{Context and Output Limits}
\label{app:cr-context}

\paragraph{Input construction.}
The evaluation code inserts the complete stored EVM assembly into each
decompilation request. It does not truncate or omit input contracts. Assembly
length is $39{,}012$ characters at the median, $101{,}839$ at P95, and
$131{,}993$ at the maximum. Including instructions and public signature
candidates, the largest user prompt is $136{,}311$ characters.
Table~\ref{tab:cr-context-limits} reports model-specific input-token maxima
and output caps for initial decompilation requests.

\begin{table}[htbp]
\centering

\normalsize
\caption{Recorded input-token maxima, output-token caps,
and length termination for the $600$ initial requests per model.
Maximum input tokens are provider-reported.}
\label{tab:cr-context-limits}
\begin{tabular}{@{}lrrr@{}}
\toprule
Model & Maximum input tokens & Output-token cap & Length stops \\
\midrule
Opus 4.7 & $113{,}519$ & $65{,}536$ & $39/600$ \\
GPT-5.3-Codex & $63{,}407$ & $65{,}536$ & $0/600$ \\
GLM-5 & $94{,}638$ & $32{,}768$ & $22/600$ \\
GLM-5 (instruct) & $64{,}726$ & $8{,}192$ & $6/600$ \\
\bottomrule
\end{tabular}
\end{table}

\begin{table}[!ht]
\centering

\normalsize
\caption{Repaired semantic consistency after excluding initial
length-terminated requests. The denominator is all callable entries in the retained contracts.}
\label{tab:cr-output-sensitivity}
\begin{tabular}{@{}lrrr@{}}
\toprule
Model & Contracts retained & All contracts & Excluding length stops \\
\midrule
Opus 4.7 & $561$ & $29.4\%$ & $4{,}284/13{,}747$ ($31.2\%$) \\
GPT-5.3-Codex & $600$ & $28.0\%$ & $4{,}078/14{,}553$ ($28.0\%$) \\
GLM-5 & $578$ & $20.7\%$ & $3{,}015/14{,}155$ ($21.3\%$) \\
GLM-5 (instruct) & $594$ & $17.7\%$ & $2{,}570/14{,}303$ ($18.0\%$) \\
\bottomrule
\end{tabular}
\end{table}

\paragraph{Output-limit sensitivity.}
Table~\ref{tab:cr-output-sensitivity} recomputes the repaired cumulative
semantic rate after excluding each model's initially length-terminated
requests. Opus 4.7 changes from $29.4\%$ to $31.2\%$, GLM-5 from $20.7\%$ to
$21.3\%$, and GLM-5 (instruct) from $17.7\%$ to $18.0\%$. GPT-5.3-Codex is unchanged
because none of its initial requests ended at the output limit. Note that
these are within-model sensitivity results. We do not compare the filtered
rates across models because their output limits and retained contract subsets differ.

\FloatBarrier
\section{Token Usage and Cost}
\label{app:pricing}
Tables~\ref{tab:model-pricing} and~\ref{tab:model-costs} report OpenRouter pricing, average token usage, and average API cost per benchmark contract. Costs are in U.S. dollars and are computed from recorded token usage and provider pricing at the time of evaluation. For $^\dagger$ rows, averages include initial decompilation and, when applicable, one zero-shot repair call.

\begin{table}[h]
\centering
\caption{OpenRouter pricing used for the cost calculation. Prices are in dollars per one million input or output tokens.}
\begin{tabular}{lcc}
\toprule
Model & Input price & Output price \\
\midrule
Opus 4.7 & $\$1.70$ & $\$25.00$ \\
GPT-5.3-Codex & $\$1.75$ & $\$14.00$ \\
GLM-5 & $\$0.60$ & $\$2.00$ \\
GLM-5 (instruct) & $\$0.60$ & $\$2.00$ \\
\bottomrule
\end{tabular}
\label{tab:model-pricing}
\end{table}

\begin{table}[h]
\centering
\caption{Average token usage and API cost by complexity stratum and overall. Token usage is average input/output tokens per contract.}
\resizebox{\linewidth}{!}{%
\begin{tabular}{lcccccccc}
\toprule
 & \multicolumn{2}{c}{Easy} & \multicolumn{2}{c}{Medium} & \multicolumn{2}{c}{Hard} & \multicolumn{2}{c}{Total} \\
\cmidrule(lr){2-3}\cmidrule(lr){4-5}\cmidrule(lr){6-7}\cmidrule(lr){8-9}
Model & Tokens & Cost & Tokens & Cost & Tokens & Cost & Tokens & Cost \\
\midrule
Opus 4.7 & $14.1$K/$17.4$K & $\$0.460$ & $35.6$K/$23.3$K & $\$0.643$ & $66.8$K/$8.0$K & $\$0.313$ & $38.8$K/$16.2$K & $\$0.472$ \\
Opus 4.7$^\dagger$ & $14.3$K/$17.6$K & $\$0.464$ & $36.4$K/$23.9$K & $\$0.661$ & $68.4$K/$9.4$K & $\$0.351$ & $39.7$K/$17.0$K & $\$0.492$ \\
GPT-5.3-Codex & $7.8$K/$13.4$K & $\$0.201$ & $19.1$K/$17.4$K & $\$0.276$ & $35.8$K/$19.8$K & $\$0.340$ & $20.9$K/$16.9$K & $\$0.273$ \\
GPT-5.3-Codex$^\dagger$ & $8.1$K/$13.6$K & $\$0.205$ & $19.7$K/$18.0$K & $\$0.287$ & $37.3$K/$21.3$K & $\$0.364$ & $21.7$K/$17.6$K & $\$0.285$ \\
GLM-5 & $8.4$K/$17.8$K & $\$0.041$ & $20.1$K/$14.6$K & $\$0.041$ & $37.4$K/$10.7$K & $\$0.044$ & $22.0$K/$14.4$K & $\$0.042$ \\
GLM-5$^\dagger$ & $8.9$K/$18.4$K & $\$0.042$ & $21.5$K/$16.9$K & $\$0.047$ & $39.8$K/$14.7$K & $\$0.053$ & $23.4$K/$16.6$K & $\$0.047$ \\
GLM-5 (instruct) & $8.1$K/$1.2$K & $\$0.007$ & $20.3$K/$2.2$K & $\$0.017$ & $38.4$K/$3.9$K & $\$0.031$ & $22.2$K/$2.4$K & $\$0.018$ \\
GLM-5 (instruct)$^\dagger$ & $8.9$K/$1.8$K & $\$0.009$ & $22.2$K/$3.7$K & $\$0.021$ & $41.4$K/$6.4$K & $\$0.038$ & $24.1$K/$3.9$K & $\$0.022$ \\
\bottomrule
\end{tabular}
}
\label{tab:model-costs}
\end{table}

One notable pattern is that Opus 4.7 is cheaper on the hard split than on the medium split, despite receiving longer inputs. This is because it produces much shorter hard-split outputs on average.

\FloatBarrier
\section{Validation and Sensitivity}
\label{app:validation-sensitivity}
\subsection{Complexity Strata and Observed Performance}
\label{app:cr-complexity}

The complexity score is a sampling instrument rather than a direct or complete
measure of LLM decompilation difficulty. We test its empirical relationship
with performance using the original 600-contract benchmark and the four
repaired model settings. Each contract contributes one observation to the
Spearman analysis in Table~\ref{tab:cr-complexity-correlations}, with its
performance averaged across models. Format is measured on the original
output because the repair protocol does not repair format failures. The later
stages use the output after the allowed compilation repair. All outcomes use
the cumulative evaluation protocol, so earlier-stage failures receive zero at
later stages. ABI F1 and semantic consistency are computed per contract for
this association analysis.

\begin{table}[htbp]

\centering
\normalsize
\setlength{\tabcolsep}{4pt}
\caption{Spearman correlations between the frozen complexity
components and across-model mean performance. Negative values indicate lower
performance as a component increases.}
\label{tab:cr-complexity-correlations}
\begin{tabular}{@{}lrrrr@{}}
\toprule
Component & Format & Compilation & ABI F1 & Semantics \\
\midrule
Bytecode scale & $-.173$ & $-.328$ & $-.346$ & $-.291$ \\
Control flow & $-.164$ & $-.326$ & $-.349$ & $-.297$ \\
Interface/source structure & $-.209$ & $-.370$ & $-.381$ & $-.315$ \\
State interaction & $-.151$ & $-.234$ & $-.271$ & $-.205$ \\
Low-level Solidity & $-.051$ & $-.177$ & $-.161$ & $-.198$ \\
Combined score & $-.183$ & $-.358$ & $-.377$ & $-.327$ \\
\bottomrule
\end{tabular}
\end{table}

The analysis shows that all five components are negatively correlated with compilation, ABI recovery,
and semantic consistency. Interface/source structure has the strongest
descriptive relationship with these outcomes. Format correlations are weaker,
and low-level Solidity features have no clear relationship with format
completeness. We suspect this is because current models are capable of
following output instructions and generating compilable Solidity, while the
generated code does not necessarily recover the original bytecode's logic.

We also fit a separate ordinary least-squares model for each outcome, using
one observation per model and contract. Predictors are the standardized
combined complexity score, $\log(1+\text{prompt characters})$,
$\log(1+\text{public-function count})$, and model fixed effects. For binary
format and compilation outcomes, this is a linear-probability analysis.
Standard errors are clustered by contract, and confidence intervals resample
complete contracts in 2,000 bootstrap replicates. This preserves the
dependence between different models' outcomes on the same contract.

After these adjustments, a one-standard-deviation increase in the score is
associated with decreases of 9.6 percentage points in compilation, 11.9 points
in ABI F1, and 8.9 points in semantic consistency. The corresponding 95\%
bootstrap intervals are $[-14.8,-4.7]$, $[-17.9,-6.5]$, and
$[-13.4,-4.4]$ percentage points. The format association is small and
inconclusive at $-2.2$ points, with interval $[-6.0,+1.4]$. Because prompt
length and the combined score are strongly correlated, the adjusted
coefficients describe the remaining association after accounting for these
covariates, not causal effects. The results support using the score to
construct complexity strata, without treating those strata as a complete
measure of LLM difficulty.

\subsection{Sensitivity to Public Signature Candidates}
\label{app:cr-signature-hints}

To assess how public signature candidates affect decompilation performance,
we evaluate GPT-5.3-Codex on 60 contracts from the original benchmark, with
20 from each complexity stratum. The no-hint setting removes the external
function-selector and event-signature candidate mappings. Assembly,
intrinsic selectors and event topics, model settings, and semantic checkpoints
remain unchanged. Table~\ref{tab:cr-signature-hints} compares the no-hint
outputs against the stored with-hint results on the same subset, both before
and after one-step repair.

\begin{table}[htbp]

\centering
\normalsize
\setlength{\tabcolsep}{3pt}
\caption{Sensitivity to public signature candidates on a
60-contract subset. Semantic consistency uses 1,308 callable entries. Perfect
denotes perfectly recovered contracts.}
\label{tab:cr-signature-hints}
\begin{tabular}{@{}p{.24\linewidth}rrrrr@{}}
\toprule
Setting & Format & Compile & ABI F1 & Semantics & Perfect \\
\midrule
With hints & 55/60 & 40/60 & .709 & 329/1,308 (25.2\%) & 3/60 \\
Without hints & 58/60 & 48/60 & .395 & 177/1,308 (13.5\%) & 2/60 \\
With hints + repair & 55/60 & 51/60 & .827 & 408/1,308 (31.2\%) & 3/60 \\
Without hints + repair & 58/60 & 55/60 & .430 & 203/1,308 (15.5\%) & 2/60 \\
\bottomrule
\end{tabular}
\end{table}

After repair, removing the mappings reduces ABI F1 from .827 to .430 and
semantic consistency from 31.2\% to 15.5\%, although compilation increases
from 85.0\% to 91.7\%. The results indicate that public signature candidates
substantially help interface and behavior recovery. Nonzero performance
without the supplied mappings shows that some recovery is still possible.

\subsection{Semantic-Test Effectiveness}
\label{app:cr-test-effectiveness}

\paragraph{Checkpoint composition.}
Table~\ref{tab:cr-case-composition} decomposes the original benchmark's
$227{,}383$ concrete cases over $14{,}553$ callable entries. Every entry has
ordinary deterministic-random cases. The other families are added when their
input types, extracted hints, or setup functions permit them. An entry can
appear in more than one family, so the entry and contract columns are not
additive.

\begin{table}[htbp]
\centering

\normalsize
\caption{Composition of semantic checkpoints in the original benchmark.}
\label{tab:cr-case-composition}
\begin{tabular}{@{}lrrrr@{}}
\toprule
Family & Cases & Share & Callable entries & Contracts \\
\midrule
Random & $91{,}454$ & $40.2\%$ & $14{,}553$ & $600$ \\
Boundary & $32{,}304$ & $14.2\%$ & $8{,}175$ & $586$ \\
Static/literal & $77{,}069$ & $33.9\%$ & $7{,}153$ & $572$ \\
Stateful & $26{,}556$ & $11.7\%$ & $13{,}754$ & $524$ \\
\bottomrule
\end{tabular}
\end{table}

\paragraph{Detected mismatches.}
We analyze the four repaired model outputs that reached semantic replay.
The unit of a model/checkpoint comparison is one concrete checkpoint replayed
against one model's output. A model/callable-entry pair fails when any of its
replayed checkpoints differs on at least one observable. These denominators
exclude entries that never reached replay, unlike the cumulative benchmark
scores. Table~\ref{tab:cr-mismatch-counts} gives the per-model counts.
In total, $344{,}055$ of $564{,}261$ comparisons ($61.0\%$) mismatch, affecting
$22{,}837$ of $36{,}784$ model/callable-entry pairs ($62.1\%$).

\begin{table}[htbp]
\centering

\normalsize
\caption{Detected mismatches among repaired outputs reaching semantic replay.
Each cell reports mismatching comparisons or failing entries over its replayed denominator.}
\label{tab:cr-mismatch-counts}
\begin{tabular}{@{}lrr@{}}
\toprule
Model & Checkpoint comparisons & Callable-entry pairs \\
\midrule
Opus 4.7 & $59{,}703/130{,}605$ & $4{,}214/8{,}498$ \\
GPT-5.3-Codex & $121{,}131/187{,}666$ & $8{,}149/12{,}227$ \\
GLM-5 & $82{,}791/126{,}976$ & $5{,}287/8{,}302$ \\
GLM-5 (instruct) & $80{,}430/119{,}014$ & $5{,}187/7{,}757$ \\
\midrule
Total & $344{,}055/564{,}261$ & $22{,}837/36{,}784$ \\
\bottomrule
\end{tabular}
\end{table}

\paragraph{Contribution of test families and observables.}
The boundary, static/literal, and stateful families together identify
$1{,}533$ failing model/callable-entry pairs missed by random cases. Of all
failing pairs, $815$ are detected only by stateful cases, $258$ only by
boundary cases, and $176$ only by static/literal cases. These exclusive counts
mean that removing the corresponding family would make the pair pass all
remaining checkpoints. They do not sum to $1{,}533$, because multiple
non-random families can detect the same failure. Observing more than return
values also contributes materially. There are $1{,}242$ failing pairs whose
discrepancies appear only in storage changes and $140$ whose discrepancies
appear only in event logs. Random-only or return-value-only checking would
therefore count some observed failures as consistent.

\subsection{Coverage Sensitivity}
\label{app:cr-coverage-sensitivity}

The benchmark has a code coverage of $77.8\%$. To assess
whether low-coverage contracts affect our findings, we use the recorded
per-contract coverage measurements to recompute repaired semantic consistency
after excluding contracts below progressively higher coverage thresholds.
The denominator includes all benchmarked callable entries in the retained
contracts, with zero credit for missing or unrecovered entries.
Table~\ref{tab:cr-coverage-sensitivity} shows that the frontier models continue
to outperform both GLM-5 settings across these subsets, while Opus 4.7 and
GPT-5.3-Codex remain close.

\begin{table}[htbp]
\centering

\normalsize
\caption{Repaired semantic consistency on subsets defined by different
code-coverage thresholds. All model columns report percentages.}
\label{tab:cr-coverage-sensitivity}
\begin{tabular}{@{}lrrcccc@{}}
\toprule
Coverage & Contracts & Entries & Opus 4.7 & GPT-5.3-Codex & GLM-5 & GLM-5 (instruct) \\
\midrule
All contracts & $600$ & $14{,}553$ & $29.4$ & $28.0$ & $20.7$ & $17.7$ \\
$\geq25\%$ & $441$ & $10{,}301$ & $31.0$ & $29.0$ & $20.1$ & $17.1$ \\
$\geq50\%$ & $369$ & $8{,}812$ & $30.5$ & $27.9$ & $19.9$ & $17.5$ \\
$\geq75\%$ & $257$ & $6{,}253$ & $30.0$ & $28.8$ & $21.1$ & $16.9$ \\
$\geq80\%$ & $239$ & $5{,}770$ & $29.7$ & $29.5$ & $21.4$ & $17.2$ \\
\bottomrule
\end{tabular}
\end{table}

\FloatBarrier
\section{Replication on a Second Contract Sample}
\label{app:cr-second-sample}

To test the stability of our findings across a different contract sample,
we reran all four models, including one-step compilation repair, on a second
600-contract sample with the same 200/200/200 complexity strata. The second
sample shares no contracts or selected near-duplicate clusters with the
original sample. It contains 226,138 replayable semantic test cases over
14,578 callable entries. Tables~\ref{tab:cr-second-format}
through~\ref{tab:cr-second-perfect} report the full results.
The dagger ($\dagger$) denotes the one-step repair protocol used in the
primary evaluation.

\begin{table}[htbp]

\centering
\normalsize
\setlength{\tabcolsep}{3pt}
\caption{Second-sample format completeness, shown as count
(rate). Each stratum contains 200 contracts, for a total of 600.
Highlighted cells indicate the best result in each column.}
\label{tab:cr-second-format}
\begin{tabular}{@{}p{.24\linewidth}rrrr@{}}
\toprule
Model & Easy & Medium & Hard & Total \\
\midrule
Opus 4.7 & 180 (90.0\%) & 146 (73.0\%) & 129 (64.5\%) & 455 (75.8\%) \\
GPT-5.3-Codex & \cellcolor{green!20}198 (99.0\%) & \cellcolor{green!20}199 (99.5\%) & \cellcolor{green!20}198 (99.0\%) & \cellcolor{green!20}595 (99.2\%) \\
GLM-5 & 116 (58.0\%) & 120 (60.0\%) & 135 (67.5\%) & 371 (61.8\%) \\
GLM-5 (instruct) & 184 (92.0\%) & 178 (89.0\%) & 164 (82.0\%) & 526 (87.7\%) \\
\bottomrule
\end{tabular}
\end{table}

\begin{table}[htbp]

\centering
\normalsize
\setlength{\tabcolsep}{3pt}
\caption{Second-sample compilation success, shown as count
(rate). Each stratum contains 200 contracts. Blue highlights the best
unrepaired result, and green highlights the best repaired result.}
\label{tab:cr-second-compilation}
\begin{tabular}{@{}p{.24\linewidth}rrrr@{}}
\toprule
Model & Easy & Medium & Hard & Total \\
\midrule
Opus 4.7 & \cellcolor{blue!12}159 (79.5\%) & 114 (57.0\%) & 77 (38.5\%) & 350 (58.3\%) \\
Opus 4.7$^\dagger$ & 176 (88.0\%) & 143 (71.5\%) & 119 (59.5\%) & 438 (73.0\%) \\
GPT-5.3-Codex & 151 (75.5\%) & \cellcolor{blue!12}136 (68.0\%) & \cellcolor{blue!12}124 (62.0\%) & \cellcolor{blue!12}411 (68.5\%) \\
GPT-5.3-Codex$^\dagger$ & \cellcolor{green!20}187 (93.5\%) & \cellcolor{green!20}190 (95.0\%) & \cellcolor{green!20}183 (91.5\%) & \cellcolor{green!20}560 (93.3\%) \\
GLM-5 & 81 (40.5\%) & 54 (27.0\%) & 42 (21.0\%) & 177 (29.5\%) \\
GLM-5$^\dagger$ & 104 (52.0\%) & 95 (47.5\%) & 89 (44.5\%) & 288 (48.0\%) \\
GLM-5 (instruct) & 80 (40.0\%) & 40 (20.0\%) & 27 (13.5\%) & 147 (24.5\%) \\
GLM-5 (instruct)$^\dagger$ & 157 (78.5\%) & 117 (58.5\%) & 85 (42.5\%) & 359 (59.8\%) \\
\bottomrule
\end{tabular}
\end{table}

\begin{table}[htbp]

\centering
\normalsize
\setlength{\tabcolsep}{3pt}
\caption{Second-sample cumulative ABI recovery, shown as
micro precision/recall/F1. Blue marks the best unrepaired F1 in each column,
and green marks the best repaired F1.}
\label{tab:cr-second-abi}
\begin{tabular}{@{}p{.24\linewidth}rrrr@{}}
\toprule
Model & Easy & Medium & Hard & Total \\
\midrule
Opus 4.7 & .933/.766/.841 & .922/.550/.689 & .924/.351/.509 & .925/.478/.630 \\
Opus 4.7$^\dagger$ & .933/.831/.879 & .916/.702/.795 & .923/.549/.688 & .923/.640/.756 \\
GPT-5.3-Codex & .977/.761/\colorbox{blue!12}{.855} & .970/.639/\colorbox{blue!12}{.770} & .942/.580/\colorbox{blue!12}{.718} & .957/.626/\colorbox{blue!12}{.757} \\
GPT-5.3-Codex$^\dagger$ & .973/.929/\colorbox{green!20}{.950} & .964/.929/\colorbox{green!20}{.946} & .937/.844/\colorbox{green!20}{.888} & .952/.884/\colorbox{green!20}{.917} \\
GLM-5 & .966/.369/.534 & .908/.255/.398 & .876/.169/.283 & .908/.227/.363 \\
GLM-5$^\dagger$ & .940/.459/.617 & .906/.447/.599 & .880/.359/.510 & .899/.402/.556 \\
GLM-5 (instruct) & .931/.361/.520 & .890/.178/.297 & .798/.115/.202 & .867/.173/.288 \\
GLM-5 (instruct)$^\dagger$ & .882/.734/.801 & .852/.536/.658 & .848/.379/.524 & .857/.483/.618 \\
\bottomrule
\end{tabular}
\end{table}

\FloatBarrier
The second sample reproduces the main findings. Repaired GPT-5.3-Codex again leads
compilation and ABI recovery at 93.3\% and .917 F1, while repaired Opus 4.7 has
the highest semantic consistency at 32.0\%. One-step repair improves the
compilation, ABI, and semantic results of every model, but full recovery
remains rare. Repaired Opus 4.7 and GPT-5.3-Codex each perfectly recover only 36 of 600
contracts. Semantic consistency decreases from easy to hard for every model,
and both frontier models remain ahead of both GLM-5 settings on repaired
compilation, ABI, and semantic metrics. These results support the stability
of the broad findings across samples, without requiring every numerical
result or model ordering to remain unchanged.

\begin{table}[!ht]

\centering
\normalsize
\setlength{\tabcolsep}{3pt}
\caption{Second-sample semantically consistent callable
entries, shown as count (rate). The easy, medium, hard, and total denominators
are 2,261, 4,623, 7,694, and 14,578.}
\label{tab:cr-second-semantics}
\begin{tabular}{@{}p{.24\linewidth}rrrr@{}}
\toprule
Model & Easy & Medium & Hard & Total \\
\midrule
Opus 4.7 & \cellcolor{blue!12}810 (35.8\%) & \cellcolor{blue!12}1,366 (29.5\%) & 1,216 (15.8\%) & \cellcolor{blue!12}3,392 (23.3\%) \\
Opus 4.7$^\dagger$ & \cellcolor{green!20}889 (39.3\%) & \cellcolor{green!20}1,725 (37.3\%) & 2,047 (26.6\%) & \cellcolor{green!20}4,661 (32.0\%) \\
GPT-5.3-Codex & 654 (28.9\%) & 886 (19.2\%) & \cellcolor{blue!12}1,416 (18.4\%) & 2,956 (20.3\%) \\
GPT-5.3-Codex$^\dagger$ & 822 (36.4\%) & 1,453 (31.4\%) & \cellcolor{green!20}2,096 (27.2\%) & 4,371 (30.0\%) \\
GLM-5 & 346 (15.3\%) & 512 (11.1\%) & 413 (5.4\%) & 1,271 (8.7\%) \\
GLM-5$^\dagger$ & 440 (19.5\%) & 880 (19.0\%) & 929 (12.1\%) & 2,249 (15.4\%) \\
GLM-5 (instruct) & 305 (13.5\%) & 222 (4.8\%) & 242 (3.1\%) & 769 (5.3\%) \\
GLM-5 (instruct)$^\dagger$ & 611 (27.0\%) & 721 (15.6\%) & 809 (10.5\%) & 2,141 (14.7\%) \\
\bottomrule
\end{tabular}
\end{table}

\begin{table}[!ht]

\centering
\normalsize
\setlength{\tabcolsep}{3pt}
\caption{Second-sample perfectly recovered contracts, shown
as count (rate). Each stratum contains 200 contracts. Ties are highlighted.}
\label{tab:cr-second-perfect}
\begin{tabular}{@{}p{.24\linewidth}rrrr@{}}
\toprule
Model & Easy & Medium & Hard & Total \\
\midrule
Opus 4.7 & 22 (11.0\%) & \cellcolor{blue!12}8 (4.0\%) & \cellcolor{blue!12}1 (0.5\%) & \cellcolor{blue!12}31 (5.2\%) \\
Opus 4.7$^\dagger$ & 25 (12.5\%) & \cellcolor{green!20}10 (5.0\%) & \cellcolor{green!20}1 (0.5\%) & \cellcolor{green!20}36 (6.0\%) \\
GPT-5.3-Codex & \cellcolor{blue!12}26 (13.0\%) & 4 (2.0\%) & 0 (0.0\%) & 30 (5.0\%) \\
GPT-5.3-Codex$^\dagger$ & \cellcolor{green!20}30 (15.0\%) & 6 (3.0\%) & 0 (0.0\%) & \cellcolor{green!20}36 (6.0\%) \\
GLM-5 & 10 (5.0\%) & 1 (0.5\%) & 0 (0.0\%) & 11 (1.8\%) \\
GLM-5$^\dagger$ & 13 (6.5\%) & 2 (1.0\%) & \cellcolor{green!20}1 (0.5\%) & 16 (2.7\%) \\
GLM-5 (instruct) & 1 (0.5\%) & 0 (0.0\%) & 0 (0.0\%) & 1 (0.2\%) \\
GLM-5 (instruct)$^\dagger$ & 3 (1.5\%) & 0 (0.0\%) & 0 (0.0\%) & 3 (0.5\%) \\
\bottomrule
\end{tabular}
\end{table}

\FloatBarrier
\section{Broader Impact and Responsible Release}
\label{app:broader-impact}
SCDBench is intended to support reproducible evaluation, security auditing, and transparency for deployed smart contracts. Better decompilation can also lower the effort required to inspect deployed contracts, which may support legitimate audits but could also assist malicious reverse engineering. We mitigate this risk by using only source-verified contracts whose bytecode and source code are already public, by releasing evaluation artifacts rather than an exploit-generation system, and by documenting semantic replay as an evaluation tool rather than a proof of correctness. The released dataset does not contain private keys, private transactions, or non-public contract sources.

\end{document}

%% file: figures/abi_metric_distribution_by_difficulty.tex
\begin{tikzpicture}[x=1cm,y=1cm,font=\fontsize{7.0pt}{7.8pt}\selectfont]
\definecolor{abiZero}{HTML}{D9D9D9}
\definecolor{abiLow}{HTML}{F4B6B6}
\definecolor{abiMid}{HTML}{FFD58A}
\definecolor{abiHigh}{HTML}{A8D5FF}
\definecolor{abiPerfect}{HTML}{7BC87C}
\definecolor{abiBest}{HTML}{E6F4EA}
\node[anchor=center,font=\bfseries] at (4.150,0.340) {Easy};
\node[anchor=center] at (4.150,0.110) {200 contracts};
\node[anchor=center,font=\bfseries] at (6.970,0.340) {Medium};
\node[anchor=center] at (6.970,0.110) {200 contracts};
\node[anchor=center,font=\bfseries] at (9.790,0.340) {Hard};
\node[anchor=center] at (9.790,0.110) {200 contracts};
\node[anchor=center,font=\bfseries] at (12.610,0.340) {Total};
\node[anchor=center] at (12.610,0.110) {600 contracts};
\node[anchor=west,font=\bfseries] at (0,0.020) {Precision};
\node[anchor=east] at (2.950,-0.275) {Opus 4.7};
\fill[abiZero] (3.150,-0.360) rectangle (3.690,-0.190);
\fill[abiLow] (3.690,-0.360) rectangle (3.730,-0.190);
\fill[abiMid] (3.730,-0.360) rectangle (3.990,-0.190);
\fill[abiHigh] (3.990,-0.360) rectangle (4.100,-0.190);
\fill[abiPerfect] (4.100,-0.360) rectangle (5.150,-0.190);
\node[anchor=west] at (5.205,-0.275) {105};
\fill[abiZero] (5.970,-0.360) rectangle (6.850,-0.190);
\fill[abiLow] (6.850,-0.360) rectangle (6.860,-0.190);
\fill[abiMid] (6.860,-0.360) rectangle (7.020,-0.190);
\fill[abiHigh] (7.020,-0.360) rectangle (7.290,-0.190);
\fill[abiPerfect] (7.290,-0.360) rectangle (7.970,-0.190);
\node[anchor=west] at (8.025,-0.275) {68};
\fill[abiZero] (8.790,-0.360) rectangle (10.010,-0.190);
\fill[abiLow] (10.010,-0.360) rectangle (10.050,-0.190);
\fill[abiMid] (10.050,-0.360) rectangle (10.190,-0.190);
\fill[abiHigh] (10.190,-0.360) rectangle (10.470,-0.190);
\fill[abiPerfect] (10.470,-0.360) rectangle (10.790,-0.190);
\node[anchor=west] at (10.845,-0.275) {32};
\fill[abiZero] (11.610,-0.360) rectangle (12.490,-0.190);
\fill[abiLow] (12.490,-0.360) rectangle (12.520,-0.190);
\fill[abiMid] (12.520,-0.360) rectangle (12.707,-0.190);
\fill[abiHigh] (12.707,-0.360) rectangle (12.927,-0.190);
\fill[abiPerfect] (12.927,-0.360) rectangle (13.610,-0.190);
\node[anchor=west] at (13.665,-0.275) {205};
\node[anchor=east] at (2.950,-0.635) {Opus 4.7\ensuremath{^{\dagger}}};
\fill[abiZero] (3.150,-0.720) rectangle (3.540,-0.550);
\fill[abiLow] (3.540,-0.720) rectangle (3.590,-0.550);
\fill[abiMid] (3.590,-0.720) rectangle (3.860,-0.550);
\fill[abiHigh] (3.860,-0.720) rectangle (3.980,-0.550);
\fill[abiPerfect] (3.980,-0.720) rectangle (5.150,-0.550);
\node[anchor=west] at (5.205,-0.635) {117};
\fill[abiZero] (5.970,-0.720) rectangle (6.580,-0.550);
\fill[abiLow] (6.580,-0.720) rectangle (6.600,-0.550);
\fill[abiMid] (6.600,-0.720) rectangle (6.850,-0.550);
\fill[abiHigh] (6.850,-0.720) rectangle (7.170,-0.550);
\fill[abiPerfect] (7.170,-0.720) rectangle (7.970,-0.550);
\node[anchor=west] at (8.025,-0.635) {80};
\fill[abiZero] (8.790,-0.720) rectangle (9.690,-0.550);
\fill[abiLow] (9.690,-0.720) rectangle (9.740,-0.550);
\fill[abiMid] (9.740,-0.720) rectangle (9.970,-0.550);
\fill[abiHigh] (9.970,-0.720) rectangle (10.370,-0.550);
\fill[abiPerfect] (10.370,-0.720) rectangle (10.790,-0.550);
\node[anchor=west] at (10.845,-0.635) {42};
\fill[abiZero] (11.610,-0.720) rectangle (12.243,-0.550);
\fill[abiLow] (12.243,-0.720) rectangle (12.283,-0.550);
\fill[abiMid] (12.283,-0.720) rectangle (12.533,-0.550);
\fill[abiHigh] (12.533,-0.720) rectangle (12.813,-0.550);
\fill[abiPerfect] (12.813,-0.720) rectangle (13.610,-0.550);
\node[anchor=west] at (13.665,-0.635) {239};
\node[anchor=east] at (2.950,-0.995) {GPT-5.3-Codex};
\fill[abiZero] (3.150,-1.080) rectangle (3.550,-0.910);
\fill[abiLow] (3.550,-1.080) rectangle (3.610,-0.910);
\fill[abiMid] (3.610,-1.080) rectangle (3.710,-0.910);
\fill[abiHigh] (3.710,-1.080) rectangle (3.770,-0.910);
\fill[abiPerfect] (3.770,-1.080) rectangle (5.150,-0.910);
\node[anchor=west] at (5.205,-0.995) {138};
\fill[abiZero] (5.970,-1.080) rectangle (6.550,-0.910);
\fill[abiLow] (6.550,-1.080) rectangle (6.560,-0.910);
\fill[abiMid] (6.560,-1.080) rectangle (6.670,-0.910);
\fill[abiHigh] (6.670,-1.080) rectangle (6.850,-0.910);
\fill[abiPerfect] (6.850,-1.080) rectangle (7.970,-0.910);
\node[anchor=west] at (8.025,-0.995) {112};
\fill[abiZero] (8.790,-1.080) rectangle (9.640,-0.910);
\fill[abiLow] (9.640,-1.080) rectangle (9.680,-0.910);
\fill[abiMid] (9.680,-1.080) rectangle (9.820,-0.910);
\fill[abiHigh] (9.820,-1.080) rectangle (10.010,-0.910);
\fill[abiPerfect] (10.010,-1.080) rectangle (10.790,-0.910);
\node[anchor=west] at (10.845,-0.995) {78};
\fill[abiZero] (11.610,-1.080) rectangle (12.220,-0.910);
\fill[abiLow] (12.220,-1.080) rectangle (12.257,-0.910);
\fill[abiMid] (12.257,-1.080) rectangle (12.373,-0.910);
\fill[abiHigh] (12.373,-1.080) rectangle (12.517,-0.910);
\fill[abiPerfect] (12.517,-1.080) rectangle (13.610,-0.910);
\node[anchor=west] at (13.665,-0.995) {328};
\fill[abiBest] (0.000,-1.495) rectangle (14.330,-1.215);
\node[anchor=east,font=\bfseries] at (2.950,-1.355) {GPT-5.3-Codex\ensuremath{^{\dagger}}};
\fill[abiZero] (3.150,-1.440) rectangle (3.260,-1.270);
\fill[abiLow] (3.260,-1.440) rectangle (3.320,-1.270);
\fill[abiMid] (3.320,-1.440) rectangle (3.450,-1.270);
\fill[abiHigh] (3.450,-1.440) rectangle (3.510,-1.270);
\fill[abiPerfect] (3.510,-1.440) rectangle (5.150,-1.270);
\node[anchor=west] at (5.205,-1.355) {164};
\fill[abiZero] (5.970,-1.440) rectangle (6.190,-1.270);
\fill[abiLow] (6.190,-1.440) rectangle (6.200,-1.270);
\fill[abiMid] (6.200,-1.440) rectangle (6.370,-1.270);
\fill[abiHigh] (6.370,-1.440) rectangle (6.570,-1.270);
\fill[abiPerfect] (6.570,-1.440) rectangle (7.970,-1.270);
\node[anchor=west] at (8.025,-1.355) {140};
\fill[abiZero] (8.790,-1.440) rectangle (9.100,-1.270);
\fill[abiLow] (9.100,-1.440) rectangle (9.180,-1.270);
\fill[abiMid] (9.180,-1.440) rectangle (9.420,-1.270);
\fill[abiHigh] (9.420,-1.440) rectangle (9.700,-1.270);
\fill[abiPerfect] (9.700,-1.440) rectangle (10.790,-1.270);
\node[anchor=west] at (10.845,-1.355) {109};
\fill[abiZero] (11.610,-1.440) rectangle (11.823,-1.270);
\fill[abiLow] (11.823,-1.440) rectangle (11.873,-1.270);
\fill[abiMid] (11.873,-1.440) rectangle (12.053,-1.270);
\fill[abiHigh] (12.053,-1.440) rectangle (12.233,-1.270);
\fill[abiPerfect] (12.233,-1.440) rectangle (13.610,-1.270);
\node[anchor=west] at (13.665,-1.355) {413};
\node[anchor=east] at (2.950,-1.715) {GLM-5};
\fill[abiZero] (3.150,-1.800) rectangle (4.060,-1.630);
\fill[abiLow] (4.060,-1.800) rectangle (4.090,-1.630);
\fill[abiMid] (4.090,-1.800) rectangle (4.280,-1.630);
\fill[abiHigh] (4.280,-1.800) rectangle (4.370,-1.630);
\fill[abiPerfect] (4.370,-1.800) rectangle (5.150,-1.630);
\node[anchor=west] at (5.205,-1.715) {78};
\fill[abiZero] (5.970,-1.800) rectangle (7.300,-1.630);
\fill[abiLow] (7.300,-1.800) rectangle (7.310,-1.630);
\fill[abiMid] (7.310,-1.800) rectangle (7.440,-1.630);
\fill[abiHigh] (7.440,-1.800) rectangle (7.610,-1.630);
\fill[abiPerfect] (7.610,-1.800) rectangle (7.970,-1.630);
\node[anchor=west] at (8.025,-1.715) {36};
\fill[abiZero] (8.790,-1.800) rectangle (10.370,-1.630);
\fill[abiLow] (10.370,-1.800) rectangle (10.380,-1.630);
\fill[abiMid] (10.380,-1.800) rectangle (10.490,-1.630);
\fill[abiHigh] (10.490,-1.800) rectangle (10.600,-1.630);
\fill[abiPerfect] (10.600,-1.800) rectangle (10.790,-1.630);
\node[anchor=west] at (10.845,-1.715) {19};
\fill[abiZero] (11.610,-1.800) rectangle (12.883,-1.630);
\fill[abiLow] (12.883,-1.800) rectangle (12.900,-1.630);
\fill[abiMid] (12.900,-1.800) rectangle (13.043,-1.630);
\fill[abiHigh] (13.043,-1.800) rectangle (13.167,-1.630);
\fill[abiPerfect] (13.167,-1.800) rectangle (13.610,-1.630);
\node[anchor=west] at (13.665,-1.715) {133};
\node[anchor=east] at (2.950,-2.075) {GLM-5\ensuremath{^{\dagger}}};
\fill[abiZero] (3.150,-2.160) rectangle (3.660,-1.990);
\fill[abiLow] (3.660,-2.160) rectangle (3.710,-1.990);
\fill[abiMid] (3.710,-2.160) rectangle (4.000,-1.990);
\fill[abiHigh] (4.000,-2.160) rectangle (4.110,-1.990);
\fill[abiPerfect] (4.110,-2.160) rectangle (5.150,-1.990);
\node[anchor=west] at (5.205,-2.075) {104};
\fill[abiZero] (5.970,-2.160) rectangle (6.600,-1.990);
\fill[abiLow] (6.600,-2.160) rectangle (6.650,-1.990);
\fill[abiMid] (6.650,-2.160) rectangle (7.020,-1.990);
\fill[abiHigh] (7.020,-2.160) rectangle (7.410,-1.990);
\fill[abiPerfect] (7.410,-2.160) rectangle (7.970,-1.990);
\node[anchor=west] at (8.025,-2.075) {56};
\fill[abiZero] (8.790,-2.160) rectangle (9.680,-1.990);
\fill[abiLow] (9.680,-2.160) rectangle (9.720,-1.990);
\fill[abiMid] (9.720,-2.160) rectangle (10.100,-1.990);
\fill[abiHigh] (10.100,-2.160) rectangle (10.450,-1.990);
\fill[abiPerfect] (10.450,-2.160) rectangle (10.790,-1.990);
\node[anchor=west] at (10.845,-2.075) {34};
\fill[abiZero] (11.610,-2.160) rectangle (12.287,-1.990);
\fill[abiLow] (12.287,-2.160) rectangle (12.333,-1.990);
\fill[abiMid] (12.333,-2.160) rectangle (12.680,-1.990);
\fill[abiHigh] (12.680,-2.160) rectangle (12.963,-1.990);
\fill[abiPerfect] (12.963,-2.160) rectangle (13.610,-1.990);
\node[anchor=west] at (13.665,-2.075) {194};
\node[anchor=east] at (2.950,-2.435) {GLM-5 (instruct)};
\fill[abiZero] (3.150,-2.520) rectangle (4.430,-2.350);
\fill[abiLow] (4.430,-2.520) rectangle (4.460,-2.350);
\fill[abiMid] (4.460,-2.520) rectangle (4.620,-2.350);
\fill[abiHigh] (4.620,-2.520) rectangle (4.680,-2.350);
\fill[abiPerfect] (4.680,-2.520) rectangle (5.150,-2.350);
\node[anchor=west] at (5.205,-2.435) {47};
\fill[abiZero] (5.970,-2.520) rectangle (7.490,-2.350);
\fill[abiLow] (7.490,-2.520) rectangle (7.520,-2.350);
\fill[abiMid] (7.520,-2.520) rectangle (7.670,-2.350);
\fill[abiHigh] (7.670,-2.520) rectangle (7.780,-2.350);
\fill[abiPerfect] (7.780,-2.520) rectangle (7.970,-2.350);
\node[anchor=west] at (8.025,-2.435) {19};
\fill[abiZero] (8.790,-2.520) rectangle (10.590,-2.350);
\fill[abiLow] (10.590,-2.520) rectangle (10.590,-2.350);
\fill[abiMid] (10.590,-2.520) rectangle (10.680,-2.350);
\fill[abiHigh] (10.680,-2.520) rectangle (10.740,-2.350);
\fill[abiPerfect] (10.740,-2.520) rectangle (10.790,-2.350);
\node[anchor=west] at (10.845,-2.435) {5};
\fill[abiZero] (11.610,-2.520) rectangle (13.143,-2.350);
\fill[abiLow] (13.143,-2.520) rectangle (13.163,-2.350);
\fill[abiMid] (13.163,-2.520) rectangle (13.297,-2.350);
\fill[abiHigh] (13.297,-2.520) rectangle (13.373,-2.350);
\fill[abiPerfect] (13.373,-2.520) rectangle (13.610,-2.350);
\node[anchor=west] at (13.665,-2.435) {71};
\node[anchor=east] at (2.950,-2.795) {GLM-5 (instruct)\ensuremath{^{\dagger}}};
\fill[abiZero] (3.150,-2.880) rectangle (3.650,-2.710);
\fill[abiLow] (3.650,-2.880) rectangle (3.720,-2.710);
\fill[abiMid] (3.720,-2.880) rectangle (4.130,-2.710);
\fill[abiHigh] (4.130,-2.880) rectangle (4.250,-2.710);
\fill[abiPerfect] (4.250,-2.880) rectangle (5.150,-2.710);
\node[anchor=west] at (5.205,-2.795) {90};
\fill[abiZero] (5.970,-2.880) rectangle (6.670,-2.710);
\fill[abiLow] (6.670,-2.880) rectangle (6.710,-2.710);
\fill[abiMid] (6.710,-2.880) rectangle (7.160,-2.710);
\fill[abiHigh] (7.160,-2.880) rectangle (7.420,-2.710);
\fill[abiPerfect] (7.420,-2.880) rectangle (7.970,-2.710);
\node[anchor=west] at (8.025,-2.795) {55};
\fill[abiZero] (8.790,-2.880) rectangle (9.790,-2.710);
\fill[abiLow] (9.790,-2.880) rectangle (9.820,-2.710);
\fill[abiMid] (9.820,-2.880) rectangle (10.350,-2.710);
\fill[abiHigh] (10.350,-2.880) rectangle (10.630,-2.710);
\fill[abiPerfect] (10.630,-2.880) rectangle (10.790,-2.710);
\node[anchor=west] at (10.845,-2.795) {16};
\fill[abiZero] (11.610,-2.880) rectangle (12.343,-2.710);
\fill[abiLow] (12.343,-2.880) rectangle (12.390,-2.710);
\fill[abiMid] (12.390,-2.880) rectangle (12.853,-2.710);
\fill[abiHigh] (12.853,-2.880) rectangle (13.073,-2.710);
\fill[abiPerfect] (13.073,-2.880) rectangle (13.610,-2.710);
\node[anchor=west] at (13.665,-2.795) {161};
\draw[black!25] (3.150,-3.120) -- (5.150,-3.120);
\draw[black!25] (5.970,-3.120) -- (7.970,-3.120);
\draw[black!25] (8.790,-3.120) -- (10.790,-3.120);
\draw[black!25] (11.610,-3.120) -- (13.610,-3.120);
\node[anchor=west,font=\bfseries] at (0,-3.580) {Recall};
\node[anchor=east] at (2.950,-3.875) {Opus 4.7};
\fill[abiZero] (3.150,-3.960) rectangle (3.690,-3.790);
\fill[abiLow] (3.690,-3.960) rectangle (3.720,-3.790);
\fill[abiMid] (3.720,-3.960) rectangle (3.840,-3.790);
\fill[abiHigh] (3.840,-3.960) rectangle (3.880,-3.790);
\fill[abiPerfect] (3.880,-3.960) rectangle (5.150,-3.790);
\node[anchor=west] at (5.205,-3.875) {127};
\fill[abiZero] (5.970,-3.960) rectangle (6.850,-3.790);
\fill[abiLow] (6.850,-3.960) rectangle (6.860,-3.790);
\fill[abiMid] (6.860,-3.960) rectangle (6.920,-3.790);
\fill[abiHigh] (6.920,-3.960) rectangle (7.060,-3.790);
\fill[abiPerfect] (7.060,-3.960) rectangle (7.970,-3.790);
\node[anchor=west] at (8.025,-3.875) {91};
\fill[abiZero] (8.790,-3.960) rectangle (10.010,-3.790);
\fill[abiLow] (10.010,-3.960) rectangle (10.030,-3.790);
\fill[abiMid] (10.030,-3.960) rectangle (10.090,-3.790);
\fill[abiHigh] (10.090,-3.960) rectangle (10.260,-3.790);
\fill[abiPerfect] (10.260,-3.960) rectangle (10.790,-3.790);
\node[anchor=west] at (10.845,-3.875) {53};
\fill[abiZero] (11.610,-3.960) rectangle (12.490,-3.790);
\fill[abiLow] (12.490,-3.960) rectangle (12.510,-3.790);
\fill[abiMid] (12.510,-3.960) rectangle (12.590,-3.790);
\fill[abiHigh] (12.590,-3.960) rectangle (12.707,-3.790);
\fill[abiPerfect] (12.707,-3.960) rectangle (13.610,-3.790);
\node[anchor=west] at (13.665,-3.875) {271};
\node[anchor=east] at (2.950,-4.235) {Opus 4.7\ensuremath{^{\dagger}}};
\fill[abiZero] (3.150,-4.320) rectangle (3.540,-4.150);
\fill[abiLow] (3.540,-4.320) rectangle (3.570,-4.150);
\fill[abiMid] (3.570,-4.320) rectangle (3.710,-4.150);
\fill[abiHigh] (3.710,-4.320) rectangle (3.750,-4.150);
\fill[abiPerfect] (3.750,-4.320) rectangle (5.150,-4.150);
\node[anchor=west] at (5.205,-4.235) {140};
\fill[abiZero] (5.970,-4.320) rectangle (6.580,-4.150);
\fill[abiLow] (6.580,-4.320) rectangle (6.600,-4.150);
\fill[abiMid] (6.600,-4.320) rectangle (6.710,-4.150);
\fill[abiHigh] (6.710,-4.320) rectangle (6.890,-4.150);
\fill[abiPerfect] (6.890,-4.320) rectangle (7.970,-4.150);
\node[anchor=west] at (8.025,-4.235) {108};
\fill[abiZero] (8.790,-4.320) rectangle (9.690,-4.150);
\fill[abiLow] (9.690,-4.320) rectangle (9.710,-4.150);
\fill[abiMid] (9.710,-4.320) rectangle (9.860,-4.150);
\fill[abiHigh] (9.860,-4.320) rectangle (10.070,-4.150);
\fill[abiPerfect] (10.070,-4.320) rectangle (10.790,-4.150);
\node[anchor=west] at (10.845,-4.235) {72};
\fill[abiZero] (11.610,-4.320) rectangle (12.243,-4.150);
\fill[abiLow] (12.243,-4.320) rectangle (12.267,-4.150);
\fill[abiMid] (12.267,-4.320) rectangle (12.400,-4.150);
\fill[abiHigh] (12.400,-4.320) rectangle (12.543,-4.150);
\fill[abiPerfect] (12.543,-4.320) rectangle (13.610,-4.150);
\node[anchor=west] at (13.665,-4.235) {320};
\node[anchor=east] at (2.950,-4.595) {GPT-5.3-Codex};
\fill[abiZero] (3.150,-4.680) rectangle (3.550,-4.510);
\fill[abiLow] (3.550,-4.680) rectangle (3.600,-4.510);
\fill[abiMid] (3.600,-4.680) rectangle (3.770,-4.510);
\fill[abiHigh] (3.770,-4.680) rectangle (3.830,-4.510);
\fill[abiPerfect] (3.830,-4.680) rectangle (5.150,-4.510);
\node[anchor=west] at (5.205,-4.595) {132};
\fill[abiZero] (5.970,-4.680) rectangle (6.550,-4.510);
\fill[abiLow] (6.550,-4.680) rectangle (6.560,-4.510);
\fill[abiMid] (6.560,-4.680) rectangle (6.660,-4.510);
\fill[abiHigh] (6.660,-4.680) rectangle (6.790,-4.510);
\fill[abiPerfect] (6.790,-4.680) rectangle (7.970,-4.510);
\node[anchor=west] at (8.025,-4.595) {118};
\fill[abiZero] (8.790,-4.680) rectangle (9.640,-4.510);
\fill[abiLow] (9.640,-4.680) rectangle (9.680,-4.510);
\fill[abiMid] (9.680,-4.680) rectangle (9.830,-4.510);
\fill[abiHigh] (9.830,-4.680) rectangle (9.970,-4.510);
\fill[abiPerfect] (9.970,-4.680) rectangle (10.790,-4.510);
\node[anchor=west] at (10.845,-4.595) {82};
\fill[abiZero] (11.610,-4.680) rectangle (12.220,-4.510);
\fill[abiLow] (12.220,-4.680) rectangle (12.253,-4.510);
\fill[abiMid] (12.253,-4.680) rectangle (12.393,-4.510);
\fill[abiHigh] (12.393,-4.680) rectangle (12.503,-4.510);
\fill[abiPerfect] (12.503,-4.680) rectangle (13.610,-4.510);
\node[anchor=west] at (13.665,-4.595) {332};
\fill[abiBest] (0.000,-5.095) rectangle (14.330,-4.815);
\node[anchor=east,font=\bfseries] at (2.950,-4.955) {GPT-5.3-Codex\ensuremath{^{\dagger}}};
\fill[abiZero] (3.150,-5.040) rectangle (3.260,-4.870);
\fill[abiLow] (3.260,-5.040) rectangle (3.310,-4.870);
\fill[abiMid] (3.310,-5.040) rectangle (3.490,-4.870);
\fill[abiHigh] (3.490,-5.040) rectangle (3.550,-4.870);
\fill[abiPerfect] (3.550,-5.040) rectangle (5.150,-4.870);
\node[anchor=west] at (5.205,-4.955) {160};
\fill[abiZero] (5.970,-5.040) rectangle (6.190,-4.870);
\fill[abiLow] (6.190,-5.040) rectangle (6.200,-4.870);
\fill[abiMid] (6.200,-5.040) rectangle (6.340,-4.870);
\fill[abiHigh] (6.340,-5.040) rectangle (6.500,-4.870);
\fill[abiPerfect] (6.500,-5.040) rectangle (7.970,-4.870);
\node[anchor=west] at (8.025,-4.955) {147};
\fill[abiZero] (8.790,-5.040) rectangle (9.100,-4.870);
\fill[abiLow] (9.100,-5.040) rectangle (9.160,-4.870);
\fill[abiMid] (9.160,-5.040) rectangle (9.400,-4.870);
\fill[abiHigh] (9.400,-5.040) rectangle (9.590,-4.870);
\fill[abiPerfect] (9.590,-5.040) rectangle (10.790,-4.870);
\node[anchor=west] at (10.845,-4.955) {120};
\fill[abiZero] (11.610,-5.040) rectangle (11.823,-4.870);
\fill[abiLow] (11.823,-5.040) rectangle (11.863,-4.870);
\fill[abiMid] (11.863,-5.040) rectangle (12.050,-4.870);
\fill[abiHigh] (12.050,-5.040) rectangle (12.187,-4.870);
\fill[abiPerfect] (12.187,-5.040) rectangle (13.610,-4.870);
\node[anchor=west] at (13.665,-4.955) {427};
\node[anchor=east] at (2.950,-5.315) {GLM-5};
\fill[abiZero] (3.150,-5.400) rectangle (4.060,-5.230);
\fill[abiLow] (4.060,-5.400) rectangle (4.090,-5.230);
\fill[abiMid] (4.090,-5.400) rectangle (4.220,-5.230);
\fill[abiHigh] (4.220,-5.400) rectangle (4.260,-5.230);
\fill[abiPerfect] (4.260,-5.400) rectangle (5.150,-5.230);
\node[anchor=west] at (5.205,-5.315) {89};
\fill[abiZero] (5.970,-5.400) rectangle (7.300,-5.230);
\fill[abiLow] (7.300,-5.400) rectangle (7.320,-5.230);
\fill[abiMid] (7.320,-5.400) rectangle (7.410,-5.230);
\fill[abiHigh] (7.410,-5.400) rectangle (7.560,-5.230);
\fill[abiPerfect] (7.560,-5.400) rectangle (7.970,-5.230);
\node[anchor=west] at (8.025,-5.315) {41};
\fill[abiZero] (8.790,-5.400) rectangle (10.370,-5.230);
\fill[abiLow] (10.370,-5.400) rectangle (10.390,-5.230);
\fill[abiMid] (10.390,-5.400) rectangle (10.470,-5.230);
\fill[abiHigh] (10.470,-5.400) rectangle (10.590,-5.230);
\fill[abiPerfect] (10.590,-5.400) rectangle (10.790,-5.230);
\node[anchor=west] at (10.845,-5.315) {20};
\fill[abiZero] (11.610,-5.400) rectangle (12.883,-5.230);
\fill[abiLow] (12.883,-5.400) rectangle (12.907,-5.230);
\fill[abiMid] (12.907,-5.400) rectangle (13.007,-5.230);
\fill[abiHigh] (13.007,-5.400) rectangle (13.110,-5.230);
\fill[abiPerfect] (13.110,-5.400) rectangle (13.610,-5.230);
\node[anchor=west] at (13.665,-5.315) {150};
\node[anchor=east] at (2.950,-5.675) {GLM-5\ensuremath{^{\dagger}}};
\fill[abiZero] (3.150,-5.760) rectangle (3.660,-5.590);
\fill[abiLow] (3.660,-5.760) rectangle (3.690,-5.590);
\fill[abiMid] (3.690,-5.760) rectangle (3.890,-5.590);
\fill[abiHigh] (3.890,-5.760) rectangle (3.940,-5.590);
\fill[abiPerfect] (3.940,-5.760) rectangle (5.150,-5.590);
\node[anchor=west] at (5.205,-5.675) {121};
\fill[abiZero] (5.970,-5.760) rectangle (6.600,-5.590);
\fill[abiLow] (6.600,-5.760) rectangle (6.630,-5.590);
\fill[abiMid] (6.630,-5.760) rectangle (6.920,-5.590);
\fill[abiHigh] (6.920,-5.760) rectangle (7.230,-5.590);
\fill[abiPerfect] (7.230,-5.760) rectangle (7.970,-5.590);
\node[anchor=west] at (8.025,-5.675) {74};
\fill[abiZero] (8.790,-5.760) rectangle (9.680,-5.590);
\fill[abiLow] (9.680,-5.760) rectangle (9.710,-5.590);
\fill[abiMid] (9.710,-5.760) rectangle (9.990,-5.590);
\fill[abiHigh] (9.990,-5.760) rectangle (10.360,-5.590);
\fill[abiPerfect] (10.360,-5.760) rectangle (10.790,-5.590);
\node[anchor=west] at (10.845,-5.675) {43};
\fill[abiZero] (11.610,-5.760) rectangle (12.287,-5.590);
\fill[abiLow] (12.287,-5.760) rectangle (12.317,-5.590);
\fill[abiMid] (12.317,-5.760) rectangle (12.573,-5.590);
\fill[abiHigh] (12.573,-5.760) rectangle (12.817,-5.590);
\fill[abiPerfect] (12.817,-5.760) rectangle (13.610,-5.590);
\node[anchor=west] at (13.665,-5.675) {238};
\node[anchor=east] at (2.950,-6.035) {GLM-5 (instruct)};
\fill[abiZero] (3.150,-6.120) rectangle (4.430,-5.950);
\fill[abiLow] (4.430,-6.120) rectangle (4.460,-5.950);
\fill[abiMid] (4.460,-6.120) rectangle (4.560,-5.950);
\fill[abiHigh] (4.560,-6.120) rectangle (4.610,-5.950);
\fill[abiPerfect] (4.610,-6.120) rectangle (5.150,-5.950);
\node[anchor=west] at (5.205,-6.035) {54};
\fill[abiZero] (5.970,-6.120) rectangle (7.490,-5.950);
\fill[abiLow] (7.490,-6.120) rectangle (7.520,-5.950);
\fill[abiMid] (7.520,-6.120) rectangle (7.620,-5.950);
\fill[abiHigh] (7.620,-6.120) rectangle (7.780,-5.950);
\fill[abiPerfect] (7.780,-6.120) rectangle (7.970,-5.950);
\node[anchor=west] at (8.025,-6.035) {19};
\fill[abiZero] (8.790,-6.120) rectangle (10.590,-5.950);
\fill[abiLow] (10.590,-6.120) rectangle (10.590,-5.950);
\fill[abiMid] (10.590,-6.120) rectangle (10.670,-5.950);
\fill[abiHigh] (10.670,-6.120) rectangle (10.720,-5.950);
\fill[abiPerfect] (10.720,-6.120) rectangle (10.790,-5.950);
\node[anchor=west] at (10.845,-6.035) {7};
\fill[abiZero] (11.610,-6.120) rectangle (13.143,-5.950);
\fill[abiLow] (13.143,-6.120) rectangle (13.163,-5.950);
\fill[abiMid] (13.163,-6.120) rectangle (13.257,-5.950);
\fill[abiHigh] (13.257,-6.120) rectangle (13.343,-5.950);
\fill[abiPerfect] (13.343,-6.120) rectangle (13.610,-5.950);
\node[anchor=west] at (13.665,-6.035) {80};
\node[anchor=east] at (2.950,-6.395) {GLM-5 (instruct)\ensuremath{^{\dagger}}};
\fill[abiZero] (3.150,-6.480) rectangle (3.650,-6.310);
\fill[abiLow] (3.650,-6.480) rectangle (3.690,-6.310);
\fill[abiMid] (3.690,-6.480) rectangle (3.930,-6.310);
\fill[abiHigh] (3.930,-6.480) rectangle (4.030,-6.310);
\fill[abiPerfect] (4.030,-6.480) rectangle (5.150,-6.310);
\node[anchor=west] at (5.205,-6.395) {112};
\fill[abiZero] (5.970,-6.480) rectangle (6.670,-6.310);
\fill[abiLow] (6.670,-6.480) rectangle (6.700,-6.310);
\fill[abiMid] (6.700,-6.480) rectangle (7.010,-6.310);
\fill[abiHigh] (7.010,-6.480) rectangle (7.400,-6.310);
\fill[abiPerfect] (7.400,-6.480) rectangle (7.970,-6.310);
\node[anchor=west] at (8.025,-6.395) {57};
\fill[abiZero] (8.790,-6.480) rectangle (9.790,-6.310);
\fill[abiLow] (9.790,-6.480) rectangle (9.830,-6.310);
\fill[abiMid] (9.830,-6.480) rectangle (10.140,-6.310);
\fill[abiHigh] (10.140,-6.480) rectangle (10.550,-6.310);
\fill[abiPerfect] (10.550,-6.480) rectangle (10.790,-6.310);
\node[anchor=west] at (10.845,-6.395) {24};
\fill[abiZero] (11.610,-6.480) rectangle (12.343,-6.310);
\fill[abiLow] (12.343,-6.480) rectangle (12.380,-6.310);
\fill[abiMid] (12.380,-6.480) rectangle (12.667,-6.310);
\fill[abiHigh] (12.667,-6.480) rectangle (12.967,-6.310);
\fill[abiPerfect] (12.967,-6.480) rectangle (13.610,-6.310);
\node[anchor=west] at (13.665,-6.395) {193};
\draw[black!25] (3.150,-6.720) -- (5.150,-6.720);
\draw[black!25] (5.970,-6.720) -- (7.970,-6.720);
\draw[black!25] (8.790,-6.720) -- (10.790,-6.720);
\draw[black!25] (11.610,-6.720) -- (13.610,-6.720);
\node[anchor=west,font=\bfseries] at (0,-7.180) {F1};
\node[anchor=east] at (2.950,-7.475) {Opus 4.7};
\fill[abiZero] (3.150,-7.560) rectangle (3.690,-7.390);
\fill[abiLow] (3.690,-7.560) rectangle (3.720,-7.390);
\fill[abiMid] (3.720,-7.560) rectangle (3.930,-7.390);
\fill[abiHigh] (3.930,-7.560) rectangle (4.100,-7.390);
\fill[abiPerfect] (4.100,-7.560) rectangle (5.150,-7.390);
\node[anchor=west] at (5.205,-7.475) {105};
\fill[abiZero] (5.970,-7.560) rectangle (6.850,-7.390);
\fill[abiLow] (6.850,-7.560) rectangle (6.860,-7.390);
\fill[abiMid] (6.860,-7.560) rectangle (6.960,-7.390);
\fill[abiHigh] (6.960,-7.560) rectangle (7.310,-7.390);
\fill[abiPerfect] (7.310,-7.560) rectangle (7.970,-7.390);
\node[anchor=west] at (8.025,-7.475) {66};
\fill[abiZero] (8.790,-7.560) rectangle (10.010,-7.390);
\fill[abiLow] (10.010,-7.560) rectangle (10.040,-7.390);
\fill[abiMid] (10.040,-7.560) rectangle (10.150,-7.390);
\fill[abiHigh] (10.150,-7.560) rectangle (10.490,-7.390);
\fill[abiPerfect] (10.490,-7.560) rectangle (10.790,-7.390);
\node[anchor=west] at (10.845,-7.475) {30};
\fill[abiZero] (11.610,-7.560) rectangle (12.490,-7.390);
\fill[abiLow] (12.490,-7.560) rectangle (12.513,-7.390);
\fill[abiMid] (12.513,-7.560) rectangle (12.653,-7.390);
\fill[abiHigh] (12.653,-7.560) rectangle (12.940,-7.390);
\fill[abiPerfect] (12.940,-7.560) rectangle (13.610,-7.390);
\node[anchor=west] at (13.665,-7.475) {201};
\node[anchor=east] at (2.950,-7.835) {Opus 4.7\ensuremath{^{\dagger}}};
\fill[abiZero] (3.150,-7.920) rectangle (3.540,-7.750);
\fill[abiLow] (3.540,-7.920) rectangle (3.570,-7.750);
\fill[abiMid] (3.570,-7.920) rectangle (3.800,-7.750);
\fill[abiHigh] (3.800,-7.920) rectangle (3.980,-7.750);
\fill[abiPerfect] (3.980,-7.920) rectangle (5.150,-7.750);
\node[anchor=west] at (5.205,-7.835) {117};
\fill[abiZero] (5.970,-7.920) rectangle (6.580,-7.750);
\fill[abiLow] (6.580,-7.920) rectangle (6.600,-7.750);
\fill[abiMid] (6.600,-7.920) rectangle (6.770,-7.750);
\fill[abiHigh] (6.770,-7.920) rectangle (7.200,-7.750);
\fill[abiPerfect] (7.200,-7.920) rectangle (7.970,-7.750);
\node[anchor=west] at (8.025,-7.835) {77};
\fill[abiZero] (8.790,-7.920) rectangle (9.690,-7.750);
\fill[abiLow] (9.690,-7.920) rectangle (9.730,-7.750);
\fill[abiMid] (9.730,-7.920) rectangle (9.920,-7.750);
\fill[abiHigh] (9.920,-7.920) rectangle (10.390,-7.750);
\fill[abiPerfect] (10.390,-7.920) rectangle (10.790,-7.750);
\node[anchor=west] at (10.845,-7.835) {40};
\fill[abiZero] (11.610,-7.920) rectangle (12.243,-7.750);
\fill[abiLow] (12.243,-7.920) rectangle (12.273,-7.750);
\fill[abiMid] (12.273,-7.920) rectangle (12.470,-7.750);
\fill[abiHigh] (12.470,-7.920) rectangle (12.830,-7.750);
\fill[abiPerfect] (12.830,-7.920) rectangle (13.610,-7.750);
\node[anchor=west] at (13.665,-7.835) {234};
\node[anchor=east] at (2.950,-8.195) {GPT-5.3-Codex};
\fill[abiZero] (3.150,-8.280) rectangle (3.550,-8.110);
\fill[abiLow] (3.550,-8.280) rectangle (3.600,-8.110);
\fill[abiMid] (3.600,-8.280) rectangle (3.750,-8.110);
\fill[abiHigh] (3.750,-8.280) rectangle (3.860,-8.110);
\fill[abiPerfect] (3.860,-8.280) rectangle (5.150,-8.110);
\node[anchor=west] at (5.205,-8.195) {129};
\fill[abiZero] (5.970,-8.280) rectangle (6.550,-8.110);
\fill[abiLow] (6.550,-8.280) rectangle (6.560,-8.110);
\fill[abiMid] (6.560,-8.280) rectangle (6.670,-8.110);
\fill[abiHigh] (6.670,-8.280) rectangle (6.910,-8.110);
\fill[abiPerfect] (6.910,-8.280) rectangle (7.970,-8.110);
\node[anchor=west] at (8.025,-8.195) {106};
\fill[abiZero] (8.790,-8.280) rectangle (9.640,-8.110);
\fill[abiLow] (9.640,-8.280) rectangle (9.680,-8.110);
\fill[abiMid] (9.680,-8.280) rectangle (9.830,-8.110);
\fill[abiHigh] (9.830,-8.280) rectangle (10.100,-8.110);
\fill[abiPerfect] (10.100,-8.280) rectangle (10.790,-8.110);
\node[anchor=west] at (10.845,-8.195) {69};
\fill[abiZero] (11.610,-8.280) rectangle (12.220,-8.110);
\fill[abiLow] (12.220,-8.280) rectangle (12.253,-8.110);
\fill[abiMid] (12.253,-8.280) rectangle (12.390,-8.110);
\fill[abiHigh] (12.390,-8.280) rectangle (12.597,-8.110);
\fill[abiPerfect] (12.597,-8.280) rectangle (13.610,-8.110);
\node[anchor=west] at (13.665,-8.195) {304};
\fill[abiBest] (0.000,-8.695) rectangle (14.330,-8.415);
\node[anchor=east,font=\bfseries] at (2.950,-8.555) {GPT-5.3-Codex\ensuremath{^{\dagger}}};
\fill[abiZero] (3.150,-8.640) rectangle (3.260,-8.470);
\fill[abiLow] (3.260,-8.640) rectangle (3.310,-8.470);
\fill[abiMid] (3.310,-8.640) rectangle (3.490,-8.470);
\fill[abiHigh] (3.490,-8.640) rectangle (3.600,-8.470);
\fill[abiPerfect] (3.600,-8.640) rectangle (5.150,-8.470);
\node[anchor=west] at (5.205,-8.555) {155};
\fill[abiZero] (5.970,-8.640) rectangle (6.190,-8.470);
\fill[abiLow] (6.190,-8.640) rectangle (6.200,-8.470);
\fill[abiMid] (6.200,-8.640) rectangle (6.370,-8.470);
\fill[abiHigh] (6.370,-8.640) rectangle (6.630,-8.470);
\fill[abiPerfect] (6.630,-8.640) rectangle (7.970,-8.470);
\node[anchor=west] at (8.025,-8.555) {134};
\fill[abiZero] (8.790,-8.640) rectangle (9.100,-8.470);
\fill[abiLow] (9.100,-8.640) rectangle (9.170,-8.470);
\fill[abiMid] (9.170,-8.640) rectangle (9.400,-8.470);
\fill[abiHigh] (9.400,-8.640) rectangle (9.800,-8.470);
\fill[abiPerfect] (9.800,-8.640) rectangle (10.790,-8.470);
\node[anchor=west] at (10.845,-8.555) {99};
\fill[abiZero] (11.610,-8.640) rectangle (11.823,-8.470);
\fill[abiLow] (11.823,-8.640) rectangle (11.867,-8.470);
\fill[abiMid] (11.867,-8.640) rectangle (12.060,-8.470);
\fill[abiHigh] (12.060,-8.640) rectangle (12.317,-8.470);
\fill[abiPerfect] (12.317,-8.640) rectangle (13.610,-8.470);
\node[anchor=west] at (13.665,-8.555) {388};
\node[anchor=east] at (2.950,-8.915) {GLM-5};
\fill[abiZero] (3.150,-9.000) rectangle (4.060,-8.830);
\fill[abiLow] (4.060,-9.000) rectangle (4.090,-8.830);
\fill[abiMid] (4.090,-9.000) rectangle (4.250,-8.830);
\fill[abiHigh] (4.250,-9.000) rectangle (4.390,-8.830);
\fill[abiPerfect] (4.390,-9.000) rectangle (5.150,-8.830);
\node[anchor=west] at (5.205,-8.915) {76};
\fill[abiZero] (5.970,-9.000) rectangle (7.300,-8.830);
\fill[abiLow] (7.300,-9.000) rectangle (7.310,-8.830);
\fill[abiMid] (7.310,-9.000) rectangle (7.430,-8.830);
\fill[abiHigh] (7.430,-9.000) rectangle (7.670,-8.830);
\fill[abiPerfect] (7.670,-9.000) rectangle (7.970,-8.830);
\node[anchor=west] at (8.025,-8.915) {30};
\fill[abiZero] (8.790,-9.000) rectangle (10.370,-8.830);
\fill[abiLow] (10.370,-9.000) rectangle (10.380,-8.830);
\fill[abiMid] (10.380,-9.000) rectangle (10.510,-8.830);
\fill[abiHigh] (10.510,-9.000) rectangle (10.650,-8.830);
\fill[abiPerfect] (10.650,-9.000) rectangle (10.790,-8.830);
\node[anchor=west] at (10.845,-8.915) {14};
\fill[abiZero] (11.610,-9.000) rectangle (12.883,-8.830);
\fill[abiLow] (12.883,-9.000) rectangle (12.900,-8.830);
\fill[abiMid] (12.900,-9.000) rectangle (13.037,-8.830);
\fill[abiHigh] (13.037,-9.000) rectangle (13.210,-8.830);
\fill[abiPerfect] (13.210,-9.000) rectangle (13.610,-8.830);
\node[anchor=west] at (13.665,-8.915) {120};
\node[anchor=east] at (2.950,-9.275) {GLM-5\ensuremath{^{\dagger}}};
\fill[abiZero] (3.150,-9.360) rectangle (3.660,-9.190);
\fill[abiLow] (3.660,-9.360) rectangle (3.710,-9.190);
\fill[abiMid] (3.710,-9.360) rectangle (3.950,-9.190);
\fill[abiHigh] (3.950,-9.360) rectangle (4.140,-9.190);
\fill[abiPerfect] (4.140,-9.360) rectangle (5.150,-9.190);
\node[anchor=west] at (5.205,-9.275) {101};
\fill[abiZero] (5.970,-9.360) rectangle (6.600,-9.190);
\fill[abiLow] (6.600,-9.360) rectangle (6.640,-9.190);
\fill[abiMid] (6.640,-9.360) rectangle (7.010,-9.190);
\fill[abiHigh] (7.010,-9.360) rectangle (7.490,-9.190);
\fill[abiPerfect] (7.490,-9.360) rectangle (7.970,-9.190);
\node[anchor=west] at (8.025,-9.275) {48};
\fill[abiZero] (8.790,-9.360) rectangle (9.680,-9.190);
\fill[abiLow] (9.680,-9.360) rectangle (9.720,-9.190);
\fill[abiMid] (9.720,-9.360) rectangle (10.100,-9.190);
\fill[abiHigh] (10.100,-9.360) rectangle (10.550,-9.190);
\fill[abiPerfect] (10.550,-9.360) rectangle (10.790,-9.190);
\node[anchor=west] at (10.845,-9.275) {24};
\fill[abiZero] (11.610,-9.360) rectangle (12.287,-9.190);
\fill[abiLow] (12.287,-9.360) rectangle (12.330,-9.190);
\fill[abiMid] (12.330,-9.360) rectangle (12.660,-9.190);
\fill[abiHigh] (12.660,-9.360) rectangle (13.033,-9.190);
\fill[abiPerfect] (13.033,-9.360) rectangle (13.610,-9.190);
\node[anchor=west] at (13.665,-9.275) {173};
\node[anchor=east] at (2.950,-9.635) {GLM-5 (instruct)};
\fill[abiZero] (3.150,-9.720) rectangle (4.430,-9.550);
\fill[abiLow] (4.430,-9.720) rectangle (4.460,-9.550);
\fill[abiMid] (4.460,-9.720) rectangle (4.600,-9.550);
\fill[abiHigh] (4.600,-9.720) rectangle (4.740,-9.550);
\fill[abiPerfect] (4.740,-9.720) rectangle (5.150,-9.550);
\node[anchor=west] at (5.205,-9.635) {41};
\fill[abiZero] (5.970,-9.720) rectangle (7.490,-9.550);
\fill[abiLow] (7.490,-9.720) rectangle (7.520,-9.550);
\fill[abiMid] (7.520,-9.720) rectangle (7.660,-9.550);
\fill[abiHigh] (7.660,-9.720) rectangle (7.860,-9.550);
\fill[abiPerfect] (7.860,-9.720) rectangle (7.970,-9.550);
\node[anchor=west] at (8.025,-9.635) {11};
\fill[abiZero] (8.790,-9.720) rectangle (10.590,-9.550);
\fill[abiLow] (10.590,-9.720) rectangle (10.590,-9.550);
\fill[abiMid] (10.590,-9.720) rectangle (10.690,-9.550);
\fill[abiHigh] (10.690,-9.720) rectangle (10.750,-9.550);
\fill[abiPerfect] (10.750,-9.720) rectangle (10.790,-9.550);
\node[anchor=west] at (10.845,-9.635) {4};
\fill[abiZero] (11.610,-9.720) rectangle (13.143,-9.550);
\fill[abiLow] (13.143,-9.720) rectangle (13.163,-9.550);
\fill[abiMid] (13.163,-9.720) rectangle (13.290,-9.550);
\fill[abiHigh] (13.290,-9.720) rectangle (13.423,-9.550);
\fill[abiPerfect] (13.423,-9.720) rectangle (13.610,-9.550);
\node[anchor=west] at (13.665,-9.635) {56};
\node[anchor=east] at (2.950,-9.995) {GLM-5 (instruct)\ensuremath{^{\dagger}}};
\fill[abiZero] (3.150,-10.080) rectangle (3.650,-9.910);
\fill[abiLow] (3.650,-10.080) rectangle (3.710,-9.910);
\fill[abiMid] (3.710,-10.080) rectangle (4.100,-9.910);
\fill[abiHigh] (4.100,-10.080) rectangle (4.360,-9.910);
\fill[abiPerfect] (4.360,-10.080) rectangle (5.150,-9.910);
\node[anchor=west] at (5.205,-9.995) {79};
\fill[abiZero] (5.970,-10.080) rectangle (6.670,-9.910);
\fill[abiLow] (6.670,-10.080) rectangle (6.710,-9.910);
\fill[abiMid] (6.710,-10.080) rectangle (7.100,-9.910);
\fill[abiHigh] (7.100,-10.080) rectangle (7.640,-9.910);
\fill[abiPerfect] (7.640,-10.080) rectangle (7.970,-9.910);
\node[anchor=west] at (8.025,-9.995) {33};
\fill[abiZero] (8.790,-10.080) rectangle (9.790,-9.910);
\fill[abiLow] (9.790,-10.080) rectangle (9.830,-9.910);
\fill[abiMid] (9.830,-10.080) rectangle (10.290,-9.910);
\fill[abiHigh] (10.290,-10.080) rectangle (10.700,-9.910);
\fill[abiPerfect] (10.700,-10.080) rectangle (10.790,-9.910);
\node[anchor=west] at (10.845,-9.995) {9};
\fill[abiZero] (11.610,-10.080) rectangle (12.343,-9.910);
\fill[abiLow] (12.343,-10.080) rectangle (12.390,-9.910);
\fill[abiMid] (12.390,-10.080) rectangle (12.803,-9.910);
\fill[abiHigh] (12.803,-10.080) rectangle (13.207,-9.910);
\fill[abiPerfect] (13.207,-10.080) rectangle (13.610,-9.910);
\node[anchor=west] at (13.665,-9.995) {121};
\draw[black!25] (3.150,-10.320) -- (5.150,-10.320);
\draw[black!25] (5.970,-10.320) -- (7.970,-10.320);
\draw[black!25] (8.790,-10.320) -- (10.790,-10.320);
\draw[black!25] (11.610,-10.320) -- (13.610,-10.320);
\node[anchor=east] at (3.030,-10.535) {Bins};
\fill[abiZero] (3.150,-10.620) rectangle (3.350,-10.450);
\node[anchor=west,font=\fontsize{6.3pt}{7.0pt}\selectfont] at (3.390,-10.535) {0};
\fill[abiLow] (4.500,-10.620) rectangle (4.700,-10.450);
\node[anchor=west,font=\fontsize{6.3pt}{7.0pt}\selectfont] at (4.740,-10.535) {(0,0.5)};
\fill[abiMid] (5.850,-10.620) rectangle (6.050,-10.450);
\node[anchor=west,font=\fontsize{6.3pt}{7.0pt}\selectfont] at (6.090,-10.535) {[0.5,0.9)};
\fill[abiHigh] (7.200,-10.620) rectangle (7.400,-10.450);
\node[anchor=west,font=\fontsize{6.3pt}{7.0pt}\selectfont] at (7.440,-10.535) {[0.9,1)};
\fill[abiPerfect] (8.550,-10.620) rectangle (8.750,-10.450);
\node[anchor=west,font=\fontsize{6.3pt}{7.0pt}\selectfont] at (8.790,-10.535) {1};
\node[anchor=west] at (13.665,0.110) {perfect count};
\end{tikzpicture}